\documentclass[floatfix,twocolumn,longbibliography,superscriptaddress,10]{revtex4-1}
\usepackage{graphicx}
\usepackage{longtable}
\usepackage{amsmath}
\usepackage[capitalize]{cleveref} 

\begin{document}

\title{The rise and fall of countries in the global value chains}
\author{Luiz G. A. Alves} 
\email{lgaalves@northwestern.edu}
\affiliation{Department of Chemical and Biological Engineering, Northwestern University, Evanston, IL 60208, USA}
\author{Giuseppe Mangioni}
\affiliation{Dipartimento di Ingegneria Elettrica, Elettronica e Informatica, University of Catania, Catania 95125, Italy}
\author{Francisco A. Rodrigues}
\affiliation{Institute of Mathematics and Computer Science, University of S\~ao Paulo, S\~ao Carlos, SP 13566-590, Brazil}
\author{Pietro Panzarasa}
\email{p.panzarasa@qmul.ac.uk}
\affiliation{School of Business and Management, Queen Mary University of London, London E1 4NS, UK}
\author{Yamir Moreno}
\email{yamir.moreno@gmail.com}
\affiliation{Institute for Biocomputation and Physics of Complex Systems (BIFI), University of Zaragoza, Zaragoza 50009, Spain}
\affiliation{Department of Theoretical Physics, University of Zaragoza, 50009 Zaragoza, Spain}
\affiliation{ISI Foundation, Turin 10126, Italy}
\date{\today}

\begin{abstract}
Countries participate in global value chains by engaging in backward and forward transactions connecting multiple geographically dispersed production stages. Inspired by network theory, we model global trade as a multi-layer network and study its power structure by investigating the tendency of eigenvector centrality to concentrate on a small fraction of countries, a phenomenon called localization transition. We show that the market underwent a significant structural variation in 2007 just before the global financial crisis. That year witnessed an abrupt repositioning of countries in the global value chains, and in particular a remarkable reversal of leading role between the two major economies, the US and China. We uncover the hierarchical structure of the multi-layer network based on countries' time series of eigenvector centralities, and show that trade tends to concentrate between countries with different power dynamics, yet in geographical proximity. We further investigate the contribution of individual industries to countries' economic dominance, and show that also within-industry variations in countries’ market positioning took place in 2007. Moreover, we shed light on the crucial role that domestic trade played in the geopolitical landscape leading China to overtake the US and cement its status as leading economy of the global value chains. Our study shows how the 2008 crisis can offer insights to policy-makers and governments on how to turn early structural signals of upcoming exogenous shocks into opportunities for redesigning countries’ global roles in a changing geopolitical landscape. 
\end{abstract}


\maketitle

\section*{Introduction}
One of the defining features of modern production systems is the organization of value chains into distinct stages that are geographically spread out across the entire globe and to which countries contribute in complex and non-linear ways.
Spatially disaggregated production systems result in a worldwide trade network in which companies of multiple countries from different production sectors exchange intermediary products along a multi-stage, non-linear and geographically boundless production trajectory ending with products and services directed at the final demand~\cite{grossman1989product}. In this network, strategic access to scarce resources plays a critical role in shaping international relationships between countries and across industries, and in enabling countries to secure and maintain economic prominence globally and over time~\cite{kindleberger1981dominance}. The study of how products and services flow within countries and from exporters to importers is therefore crucial to better understand how countries can secure prominent roles in the global value chains.

Recent years have witnessed a growing number of network studies concerned with the worldwide trade system~\cite{cingolani2017countries,cristelli2015heterogeneous,fagiolo2009world,formichini2019influence,garlaschelli2004fitness,he2010structure,hidalgo2009building,schweitzer2009economic,serrano2003topology}. Using simplex networks to represent countries as nodes and transactions as directed links from exporters to importers, it has been suggested that the worldwide trade network exhibits a community structure~\cite{barigozzi2011identifying,piccardi2012existence}, a heavy-tailed degree distribution~\cite{fagiolo2009world}, and small-world properties~\cite{serrano2003topology}. By extending the simplex representation to bipartite networks, in which nodes in one class can only connect to nodes in another class~\cite{newman2018networks}, researchers have focused on early signals of economic downturns~\cite{saracco2015randomizing}, the competitiveness of countries~\cite{cristelli2013measuring}, and the complexity of products~\cite{hidalgo2009building}.  

In spite of these recent advances~\cite{Amaral2004}, a number of scholars have highlighted the limitations of simplex network representations and projected networks, and in particular have emphasized how these network paradigms can poorly capture the dynamics of complex systems~\cite{battiston2014structural,aleta2019multilayer}. In contexts where the system has multiple layers interconnected with each other -- such as, for example, the case of transportation systems where fluxes of individuals traveling by bus, train, and airplane can be seen as belonging to different layers of the same interconnected network -- the multi-layer network has been shown to provide a more adequate representation~\cite{aleta2019multilayer,Bianconi2018,Boccaletti2014}. This is also the case of the international trade network, in which the structure unfolds within and across industries, and the countries are involved in multiple stages of production along the global value chains~\cite{a2018unfolding}. In the multi-layer representation, industries refer to the layers of the network, the countries are the nodes that populate every layer, and connections can be drawn from one country to another in the same industry (within-layer connections), or between different industries (cross-layer connections)~\cite{alves2019nested,a2018unfolding}. 

Over recent years, researchers have adopted a multi-layer perspective to better capture the structure and dynamics of international trade~\cite{mastrandrea2014reconstructing,lee2016strength,ghariblou2017shortest,alves2019nested,a2018unfolding,formichini2019influence}. Applications of multi-layer networks to trade include the analysis of layer-specific local constraints on international trade~\cite{mastrandrea2014reconstructing}, the study of the emergence and unfolding of cascading failures~\cite{lee2016strength}, the study of countries' influence based on betweenness centrality measures~\cite{ghariblou2017shortest}, the analysis of the nested structural organization of the worldwide trade~\cite{alves2019nested}, and the assessment of the impact of technological innovations on industrial products~\cite{formichini2019influence}, to name only a few. However, despite the increasing popularity of network approaches to global trade, relatively little attention has been paid to the formalization of countries' economic dominance from a multi-layer perspective. Traditionally, scholars and policy-makers interested in comparative assessments of countries’ global roles and competitive advantage have relied on macro-economic measures of market power based on countries' overall share of world trade~\cite{lejour2014}. These traditional aggregate measures, however, suffer from shortcomings, mainly because aggregate trade flows, on which dominance is predicated, cannot account for the increasingly widespread internationalization of production processes~\cite{antras2012,antras2013,cingolani2017countries}.

Countries’ economic dominance of global trade is intrinsically rooted in the international fragmentation of production and the resulting structural intricacies that characterize the global value chains. Production processes are typically disaggregated into various stages stretching across multiple countries so as to exploit the comparative advantages of locations. For example, production might involve intermediary stages and assembly spots located in countries that hold only a negligible share of the market of the final product. Countries may receive inputs from multiple suppliers and, in turn, contribute to the production process at multiple stages located in various other countries. Thus, the internationalization of production makes it difficult to assess countries’ role in global trade simply by using traditional aggregate values such as gross imports or exports. By contrast, a country’s economic dominance of global trade should be a function of the share of value the country brings to each production stage of the underlying global value chains and of the share of benefits obtained from the exchange of intermediate and final products~\cite{costinot2013,johnson2012}. That is, economic dominance is a multi-layer property that should quantify the salience of a country for the entire production network, and thus explicitly depend on the centrality of the other countries located at other production stages and with which the focal country is connected. 

Here, we take a step in this direction, and investigate the dynamics of dominance in the worldwide multi-layer trade network using data from the World Input-Output Database (WIOD)~\cite{timmer2015illustrated}. The WIOD is an information-rich data set including details about trade among $56$ industries and $43$ countries accounting for more than 85\% of the overall global GDP in the period from 2000 to 2014. Drawing on this data set, we take a micro perspective and investigate the dynamics of economic dominance of individual buyers and sellers in the worldwide trade multi-layer network. We show that the ranking of buyers and sellers varied over time in a non-trivial way, resulting in an abrupt and remarkable reversal of leading roles precisely before the 2008 financial crisis. Using hierarchical clustering analysis, we uncover common patterns of variation in economic dominance, and use these patterns to partition countries into  meaningful groups characterized by similar power dynamics. We then shift focus from individual countries and take a macro perspective. First, we identify a localization effect in the system using the inverse participation ratio, and examine whether variations in localization can be associated with the occurrence of exogenous shocks. Second, we shed light on the contributions of individual industries on countries' global dominance, and uncover the localization effect within each industry by computing the corresponding inverse participation ratio. Finally, we explore the role that domestic trade played in amplifying the system's power concentration and facilitating variations in the geo-political landscape before the 2008 crisis. 


\section*{Results}
\begin{figure*}[!t]
\centering
\includegraphics[width=1\textwidth]{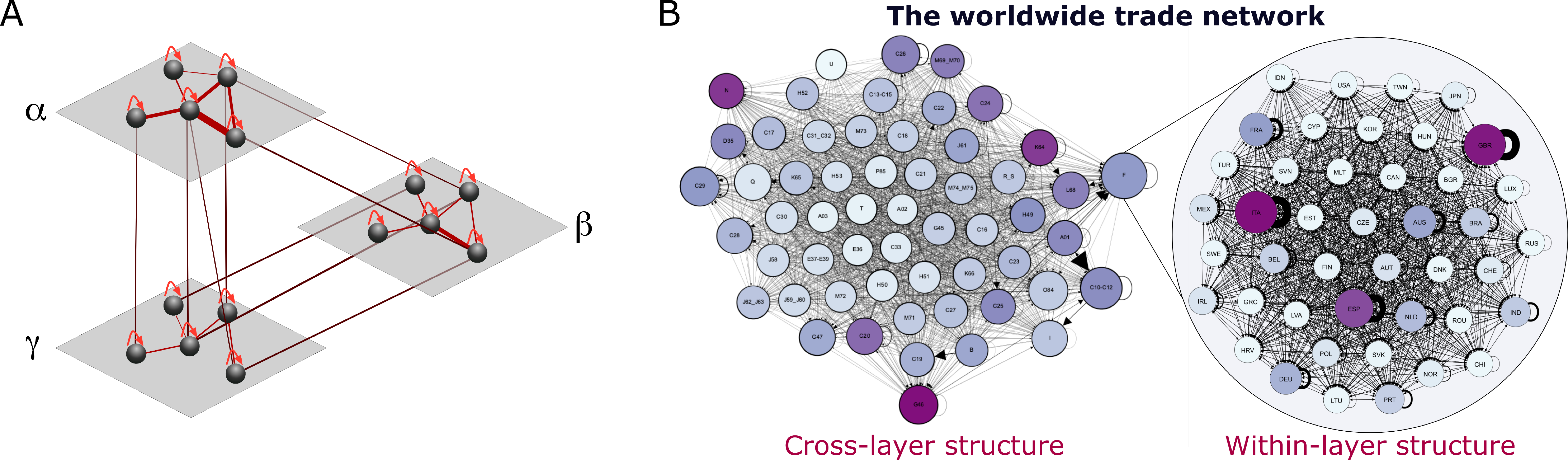}
\caption{{\bf The worldwide trade multi-layer network}. {\bf A)} Schematic representation of the multi-layer network. Each node represents a country and each layer an economic industry. The edges represent economic transactions starting from sellers and pointing to the buyers. The widths of edges are proportional to the USD value of the goods or services exchanged between the connected countries. {\bf B)} The worldwide trade network in 2000. For illustrative purposes, we show: (i) the aggregated cross-layer structure of the network, where edges refer to connections among countries from different industries; and (ii) an example of within-layer structure, where edges refer to connections between countries within the construction industry. The size of each node is proportional to the in-strength of the node, $s^{in}$, whereas the color intensity of each node is proportional to the out-strength of the node, $s^{out}$. See Tables~\ref{t:countries} and~\ref{t:activities} for the labels of the nodes.}
\label{fig:networks}
\end{figure*}  

\subsection*{The multi-layer network}

Our analysis begins with the construction of the multi-layer network using the WIOD data set (Fig~\ref{fig:networks}). The WIOD is an information-rich data set that includes details about trade among $56$ industries and $43$ countries, from 2000 to 2014. Unlike other data sets (e.g., COMTRADE), the WIOD has information about connections between different industries and different countries. Moreover, the data set enables the construction of a weighted directed multi-layer network, where each industry is represented as a layer and every layer is populated by the $43$ countries of the data set, which are connected when they trade with one another~\cite{a2018unfolding,alves2019nested}. In particular, within-layer connections refer to the exchange of products and services among countries within the same industry, whereas cross-layer connections refer to economic transactions among countries in different industries (see Fig~\ref{fig:networks} and Materials and Methods). 

\subsection*{Economic dominance of buyers and sellers}

\begin{figure*}[!t]
\centering
\includegraphics[width=1\textwidth]{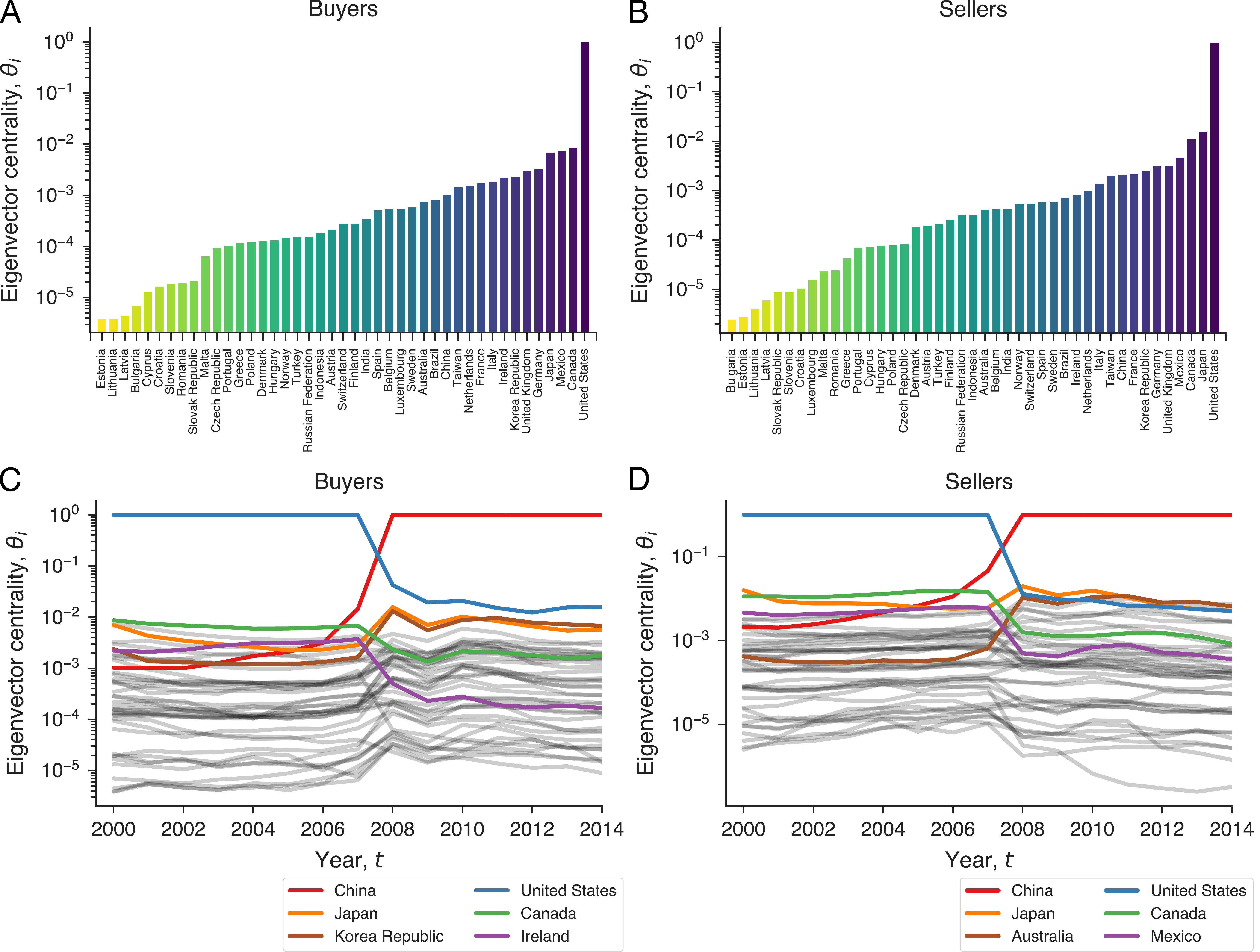}
\caption{{\bf Dynamics of economic dominance.} The eigenvector centrality of buyers and sellers in the worldwide trade multi-layer network in 2000 is shown in panels {\bf A)} and {\bf B)}, respectively. Countries are ordered by their centrality from the least central to the most central. The rankings of buyers and sellers show some similarity, as suggested by the Kendall-$\tau$ correlation coefficient ($0.82$ with a $p-$value of $<10^{-14}$). Countries that are central buyers tend to be also central sellers. The variation of eigenvector centrality over time for buyers {\bf C)} and sellers {\bf D)} shows the emergence of new winners and losers just before the 2008 financial crisis. The centralities of China and the US are shown in red and blue solid lines, respectively. Panels {\bf C)} and {\bf D)} highlight the three countries that experienced the largest positive and negative variations in eigenvector centrality between 2007 and 2008, respectively in the buyers' and sellers' market.}
\label{fig:centrality}
\end{figure*}  

In graph theory, a common measure for assessing the importance of nodes in a network is the eigenvector centrality~\cite{bonacich1987power,newman2016mathematics,newman2018networks}. The idea underpinning this measure is that a node is important to the extent that it is connected to other important nodes, and thus belongs to a chain in which importance is transmitted from one node to another along various connections~\cite{bonacich1987power}. Research has shown that under commonly occurring conditions (e.g., in networks with heavy-tail distributions) the eigenvector centrality undergoes a localization transition as a result of which most of the weight of the centrality will concentrate on a small number of nodes in the network, while the vast majority of the remaining nodes will be assigned only a vanishing fraction~\cite{sola2013eigenvector,de2015ranking,de2017disease}. This phenomenon, typically referred to as ``localization effect'', has been extensively studied in the case of simplex complex networks~\cite{goltsev2012localization}, and more recently it has been investigated in multi-layer networks~\cite{sola2013eigenvector,de2015ranking,de2017disease}. Traditionally the literature has regarded the localization effect as a drawback of the eigenvector centrality, which in turn has motivated the proposal of alternative centrality measures better able to assess the relative importance of peripheral nodes that would otherwise remain indistinguishable. Here, we take a different perspective: we shift focus from the ranking of peripheral nodes to the emergence of power structures in which a minority of nodes take on a leading role. In particular, we explore how the localization process in the multi-layer trade network can unmask important structural changes in global trade associated with the emergence of few dominant players. We also assess how these changes can result in transformations of power hierarchies in the global value chains, and ultimately affect the relative economic dominance of the leading global suppliers and destination markets.

First, we define the economic dominance of a country from a network-based perspective. More generally, a country can be seen as a global leader when it exerts control over the whole range of transactions occurring along the international global value chains. This implies that a country's global dominance should be a function of the dominance of all its trading partners at the local levels of the upstream, midstream, and downstream stages of production within and across all industries. Moreover, countries can be global leaders as both buyers and sellers. On the one hand, a country is a leading global buyer to the extent that it represents a major destination market of intermediary and finished products sold by countries that, in turn, are key destination markets of products originating from other key destination markets, and so forth. On the other, a country is a leading global supplier to the extent that it controls the sales of intermediary and finished products to countries that, in turn, are key suppliers of products to other key suppliers, and so forth.

We measure the economic dominance of buyers and sellers using a suitable adaptation of eigenvector centrality to the multi-layer network. The centrality of nodes can be measured through the tensorial formulation of the multi-layer network by calculating the spectral properties of the graph. Specifically, the corresponding formalization of the multi-layer network in the tensorial notation is rank-4 tensor $M_{i\alpha}^{j\beta}$, which encodes a directed, weighted connection between node $i$ from layer $\alpha$ to any other node $j$ in any layer $\beta$~\cite{de2013mathematical,de2015ranking,de2017disease}. Thus, we can compute the centrality of buying nodes in a given layer by calculating the leading eigentensor associated with the most positive eigenvalue~\cite{de2015ranking}. To also account for sellers, we can use the left leading eigentensor when calculating the centrality (or, equivalently, we can calculate the leading eigentensor associated with the most positive eigenvalue of the transposed supra-matrix). Thus, the node centrality can be obtained solving the following equation~\cite{de2013mathematical}:
\begin{equation}
    M_{i\alpha}^{j\beta}\Theta_{i\alpha}=\lambda \Theta_{j\beta},
\end{equation}
where $\Theta_{i\alpha}$ gives the eigenvector centrality of each node $i$ in each layer $\alpha$ when accounting for the whole interconnected multi-layer structure. The centrality of each node in the whole multi-layer network is given by aggregating over the layers the centrality of each node in each layer. Formally, this is equivalent to multiplying $\Theta_{i\alpha}$ by a rank-1 tensor ($u^\alpha$) with all components equal to 1~\cite{de2015ranking,sola2013eigenvector}, namely
\begin{equation}
    \theta_i^M=\Theta_{i\alpha}u^\alpha.
\label{centrality:multilayer}
\end{equation} 
Solving the equation above for buyers and sellers in the multi-layer network in the year 2000, the eigenvector centrality ranks the US as the dominant country both as a buyer and as a seller. The least dominant countries in 2000 are Estonia and Bulgaria, appearing at the bottom of the rankings of buyers and sellers, respectively. The values of eigenvector centrality in the multi-layer network for this year is shown in Fig.~\ref{fig:centrality}A (buyers) and  Fig.~\ref{fig:centrality}B (sellers). The rankings of buyers and sellers show some similarity, as indicated by the Kendall-$\tau$ correlation coefficient ($0.82$; $p-$value $<10^{-14}$). Countries that are prominent destination markets tend to be also prominent suppliers. The majority of countries change ranks only by a position or two, with the noticeable exception of Luxembourg that occupies position number $17$ in the ranking of buyers and position number $36$ in the ranking of sellers.  

We now focus on the dynamics of dominance in the worldwide trade multi-layer network. We computed the eigenvector centrality for each year in the period from 2000 to 2014, for both buyers (Fig.~\ref{fig:centrality}C) and sellers (Fig.~\ref{fig:centrality}D).
Interestingly, our results suggest that an abrupt reversal of economic dominance between the US and China took place precisely just before the 2008 global financial crisis. Findings also clearly indicate that many other countries changed their global roles during the crisis, and new winners and losers emerged in a reshaped geo-political landscape. In particular, in addition to the US and China, (Fig.~\ref{fig:centrality}C and  Fig.~\ref{fig:centrality}D highlight, respectively, the three buyers and the three suppliers that experienced the largest positive and negative variations in eigenvector centrality between 2007 and 2008. For example, in 2008, like China, Japan emerged as a prominent economy in global production, whereas Canada followed the same trajectory as the US, permanently losing the role occupied before the crisis as the world's second most prominent trading nation.

\subsection*{Blocks of economic dominance}

\begin{figure*}[!t]
\centering
\includegraphics[width=0.96\textwidth]{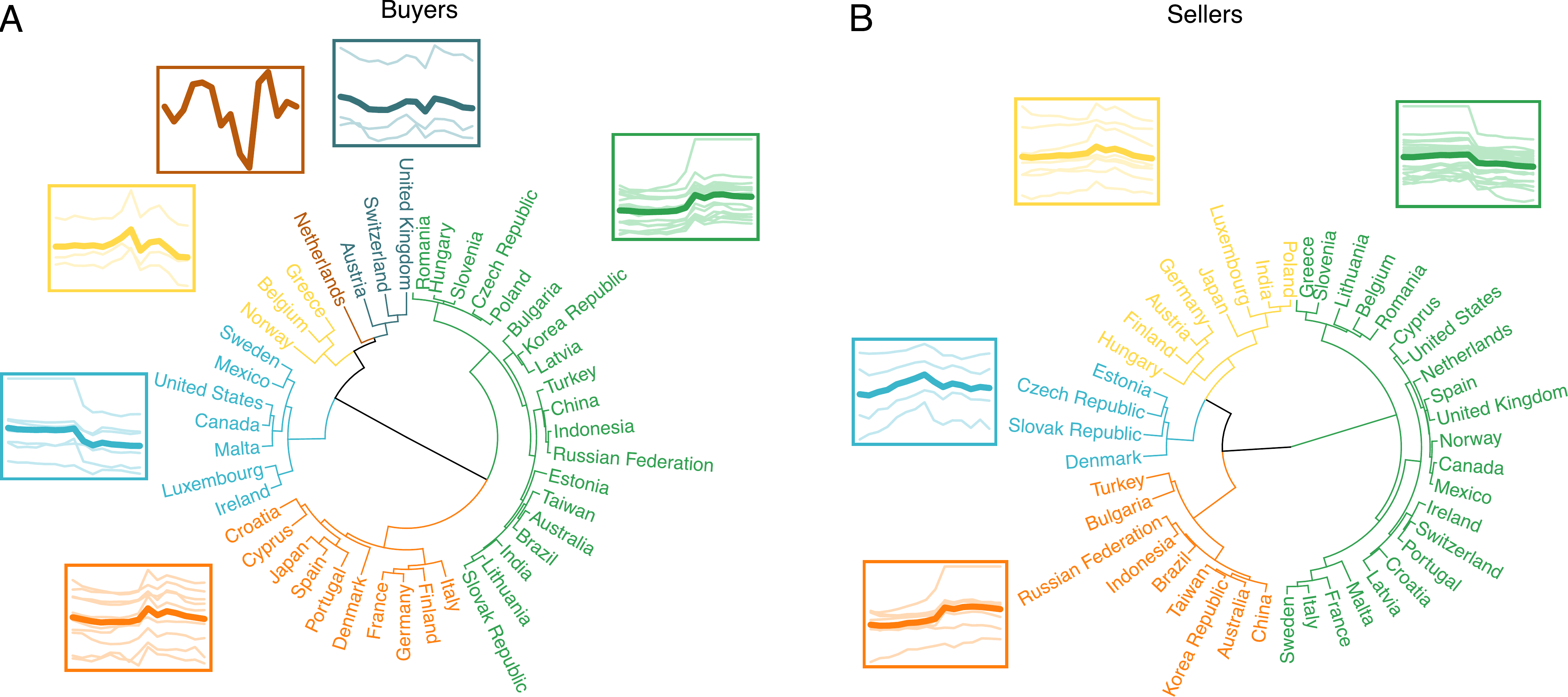}
\caption{{\bf Hierarchical structure of the worldwide trade network.} The dendrograms show the result of the hierarchical clustering based on the Pearson's correlation distance using the Ward linkage criteria for buyers (A) and sellers (B). The colors indicate the clusters obtained by cutting the dendrograms at the threshold distance that maximizes the silhouette score. The insets show the time series of the eigenvector centralities (thinner lines in light-shaded colors) of the countries belonging to each cluster (y-axis on log-scale). The thicker lines refer to the average trends of each cluster. The buyers' hierarchical structure has six clusters, whereas the sellers' hierarchical structure has only four. }
\label{fig:dendrogram}
\end{figure*}  

To further understand the emergence and evolution of groups of trading countries, we investigate the hierarchical structure of countries based on dynamics of economic dominance. By computing the Pearson's correlation distance matrix (see Materials and Methods and Fig.\ref{sfig:matrix}) and using the Ward linkage criteria~\cite{hastie2013elements,ward1963hierarchical}, we constructed a dendrogram that hierarchically clusters similar time series of eigenvector centrality. This clustering procedure recursively merges pair of clusters that minimally increase within-cluster variance. We determined the number of significant clusters by cutting the dendrogram at the threshold distance that maximizes the silhouette score~\cite{rousseeuw1987silhouettes,sigaki2019clustering}. Fig.~\ref{fig:dendrogram} shows the obtained clusters of buyers (A) and sellers (B) and the corresponding average trends of eigenvector centrality (see also the matrix plot of the correlation distance among all pairs of time series in Fig.~\ref{sfig:matrix}). To check for robustness, we also computed the modular structure of the network using a hierarchical (nested) stochastic block model, and obtained a substantial overlap with the partitioning based on the hierarchical cluster analysis (see Fig.~\ref{fig:nested_SBM}). 

Findings suggest that destination markets and suppliers can be grouped according to similar patterns of variation in the role they occupied in the economic system over time. In particular, Fig.~\ref{fig:dendrogram}A shows that buyers are grouped into six economic blocks, the largest of which (by number of members) contains $19$ countries (including Brazil, Russia, India, China). This block accounted for $20\%$ of all purchases in 2000 and overtook all the other blocks by 2007, reaching $\approx 50\%$ of all purchases in 2014 (see Fig.~\ref{fig:blocks_fraction_total}A). All the other blocks reduced their share of global purchases over time. The second largest block includes $10$ countries ($9$ European countries and Japan), and accounted for $20\%$ of all purchases in 2014. The third block, including $8$ countries (five European countries and the three North American countries, namely, Canada, the US, and Mexico), boasted the largest share of all purchases in 2000 (approximately $36\%$), but played a weaker role in 2014 ($24\%$ of global purchases). The other two clusters include three countries each (all from Europe, representing about $5\%$ of all trades, in 2014); the remaining cluster is composed of a single buyer (the Netherlands), and accounts for about $1\%$ of all purchases during the period.

Like buyers, sellers can be hierarchically grouped into clusters characterized by similar variations of economic dominance. In particular, as shown by Fig.~\ref{fig:dendrogram}B, sellers are partitioned into four economic blocks, the largest of which includes $22$ countries ($19$ from Europe and the other three from North America, namely, Canada, the US, and Mexico). This block accounted for $56\%$ and $40\%$ of all global sales, respectively in 2000 and 2014 (Fig.~\ref{fig:blocks_fraction_total}C). The second largest economic block, which includes $9$ countries (six Asian countries, Australia, Brazil, and Bulgaria), overtook the largest block only in 2013, with $43\%$ of all sales in 2014. The third economic block includes $8$ countries (six European countries and two Asian countries) and accounted for less than $2\%$ of all sales in 2014. Although the last economic block is the smallest by number of countries (only four European countries), it accounted for $15\%$ of all sales in 2014. 
 
We now shed some light on the drivers of such partitioning. First, it is interesting to examine the extent to which the grouping of countries based on patterns of variation in economic role overlaps with a more traditional partitioning based on the idea that countries are more likely to trade within than between groups~\cite{fortunato2010}. While Fig.~\ref{fig:blocks_fraction_total}B,D seems to suggest that over the years the total purchase taking place within blocks represents a fraction that varies from 65\% (for the smallest blocks) to 95\% (for the largest blocks), this is only true when domestic trade is accounted for. Indeed, as shown in Fig.~\ref{fig:blocks_fraction_international}, when the analysis is restricted only to international trade, the reverse is true. Except for the third largest block of importers (including the US, Canada, and Mexico) and the largest block of exporters (still including the three North American countries), countries tend to trade between rather than within blocks. Thus, the majority of trade occurring within blocks tends to originate from domestic trade, i.e., self-loops within and across layers (see Fig.~\ref{fig:blocks_fraction_domestic}). As suggested by~\cref{fig:importers_model1,fig:exporters_model1,fig:importers_model2,fig:exporters_model2,fig:importers_model3,fig:exporters_model3}, the economic value of the transactions within blocks does not statistically significantly differ from what would be expected by chance (at the $0.05$ significance level), except for the third largest block of importers (including the US, Canada, and Mexico) and the largest block of exporters (still including the three North American countries). Countries with similar power dynamics tend, therefore, to avoid trading with each other, and instead concentrate their transactions with partners belonging to other clusters. 

Second, the obtained groups have a geographical signature too. As suggested by Table~\ref{table:distance_buyers} and Table~\ref{table:distance_sellers}, the mean geographic distance between countries within the same block (computed using the geographical centroids of countries) is always lower (except for the largest block of buyers) than the mean distance separating countries from different blocks. Thus, spatial proximity can be regarded as a fundamental driving force underpinning the formation of clusters of trading countries characterized by comparable dynamics of economic dominance (see also the geographic mapping of countries in Fig.~\ref{fig:maps}.)

\subsection*{Inverse participation ratio}
\begin{figure*}[!t]
\centering
\includegraphics[width=1\textwidth]{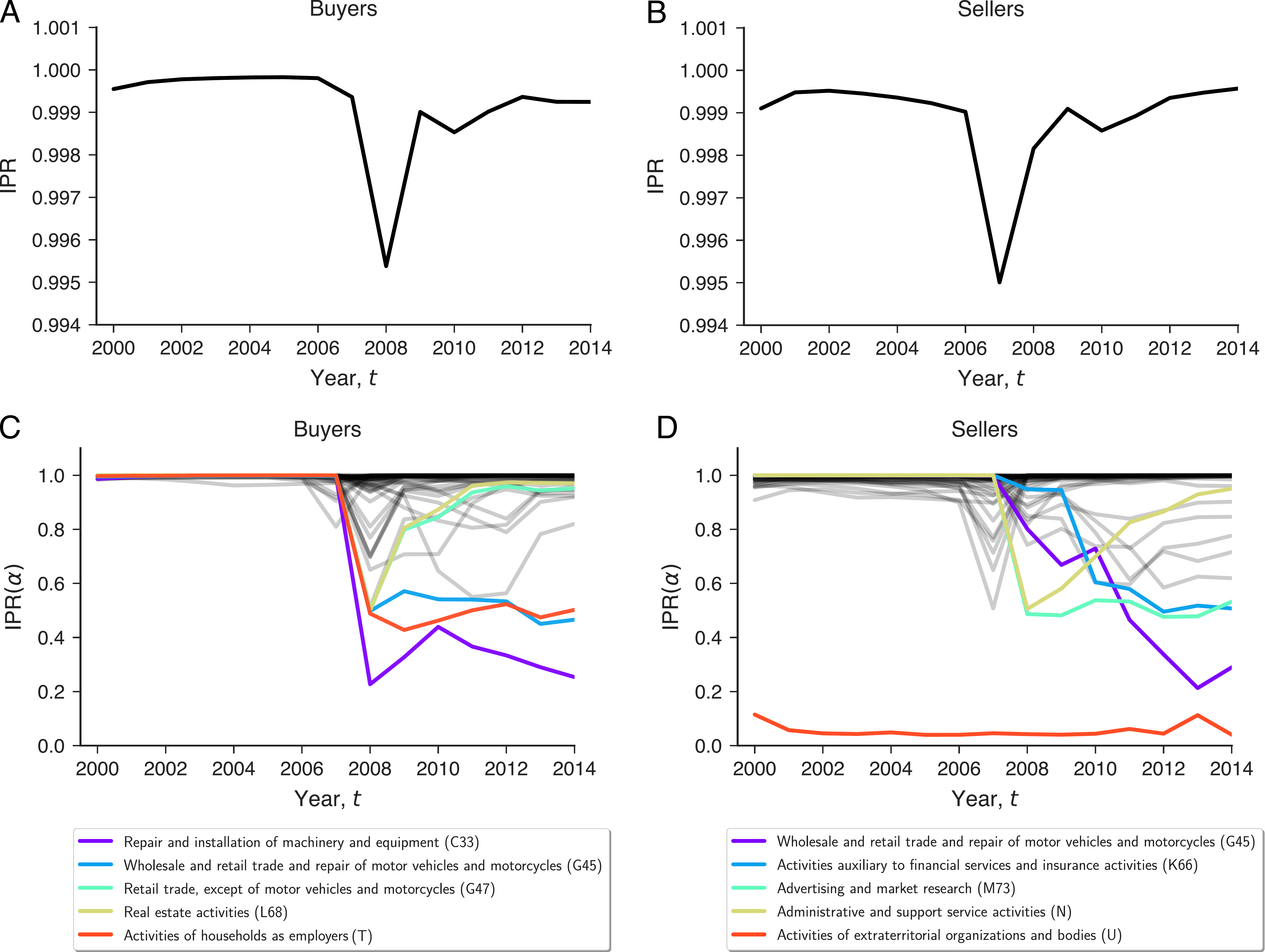}
\caption{{\bf Dynamics of the inverse participation ratio (IPR).} {\bf A-B)} The panels show a drop in the IPR during the financial crisis. This happened between 2007 and 2008 for the buyers {\bf A)} and between 2006 and 2007 for the sellers {\bf B)}. {\bf C-D)} Contribution of industries to buyers' and sellers' dominance. Contributions of an individual layer $\alpha$ to the inverse participation ratio (IPR$(\alpha)$). The colored industries are the ones that experienced the largest drop in IPR$(\alpha)$ during the observed period. Notice that most of the industries that experienced a drop in localization did not revert back to the initial localized state.}
\label{fig:ipr}
\end{figure*}

To further investigate the dynamics of eigenvector centrality, we compute the inverse participation ratio (IPR) of buyers and sellers. From network theory~\cite{martin2014localization,de2017disease}, we know that
\begin{equation}
\text{IPR}=\sum_i \theta_i^4,
\label{eq:ipr}
\end{equation}
where $\theta_i$ is the eigenvector centrality of node $i$. An IPR close to zero means that there is negligible localization effect (i.e., no economic dominance), and centrality is homogeneously distributed across the nodes, whereas an IPR close to one reflects a network where the centrality is very localized in very few nodes (i.e., the system is dominated by a minority of countries)~\cite{martin2014localization}. 

Using Eq.~\ref{eq:ipr}, we computed the IPR for the worldwide trade multi-layer network over the period from 2000 to 2014, for buyers (Fig.~\ref{fig:ipr}A) and sellers (Fig.~\ref{fig:ipr}B). Findings clearly suggest a drop in the localization effect that took place precisely before the 2008 financial crisis. Combined with the observed changes in countries' eigenvector centrality (see Fig.~\ref{centrality:multilayer}), the drop in localization indicates a more homogeneous redistribution of economic dominance across trading countries. Notice that, despite the similar shape, the drop in IPR for sellers precedes in time the one for buyers. This may suggest some linkages between global supply and destination markets: early-warning structural signals of the 2008 crisis started to appear in the sellers' market and then propagated along the global value chains to also affect the structure of the global purchasing market.  

To assess the statistical significance of the high values of IPR (i.e., to test the hypothesis that the observed values of IPR values could be generated by chance), we used a random-edge assignment to construct an ensemble of synthetic multi-layer networks, with $1,000$ realizations, and with the same topological features as the empirical networks. To produce the null model, we randomly redirected the in- or out-going edges by preserving their weights, thus also preserving the in- or out-degree distributions and the in- or out-strength distributions in the different layers, and then computed the eigenvector centrality (see Material and Methods). So constructed, this null model preserves the in- and out-degree distributions, in- and out-strength distributions, as well as the weight distribution at three levels: (i) globally, in the entire network; (ii) at the level of cross-layer connections; and (iii) locally within each layer. For each synthetic multi-layer network, we computed the IPR values for buyers and sellers to obtain a distribution of values that can then be compared with the results obtained from the empirical multi-layer networks. Assuming a 5\% false discovery rate, we can reject the hypothesis that the localization effect found in the multi-layer network can be generated by chance using a null model that has the same topological features as the real multi-layer network (see Fig.~\ref{fig:null_model}). 

\begin{figure*}[!t]
\centering
\includegraphics[width=0.75\textwidth]{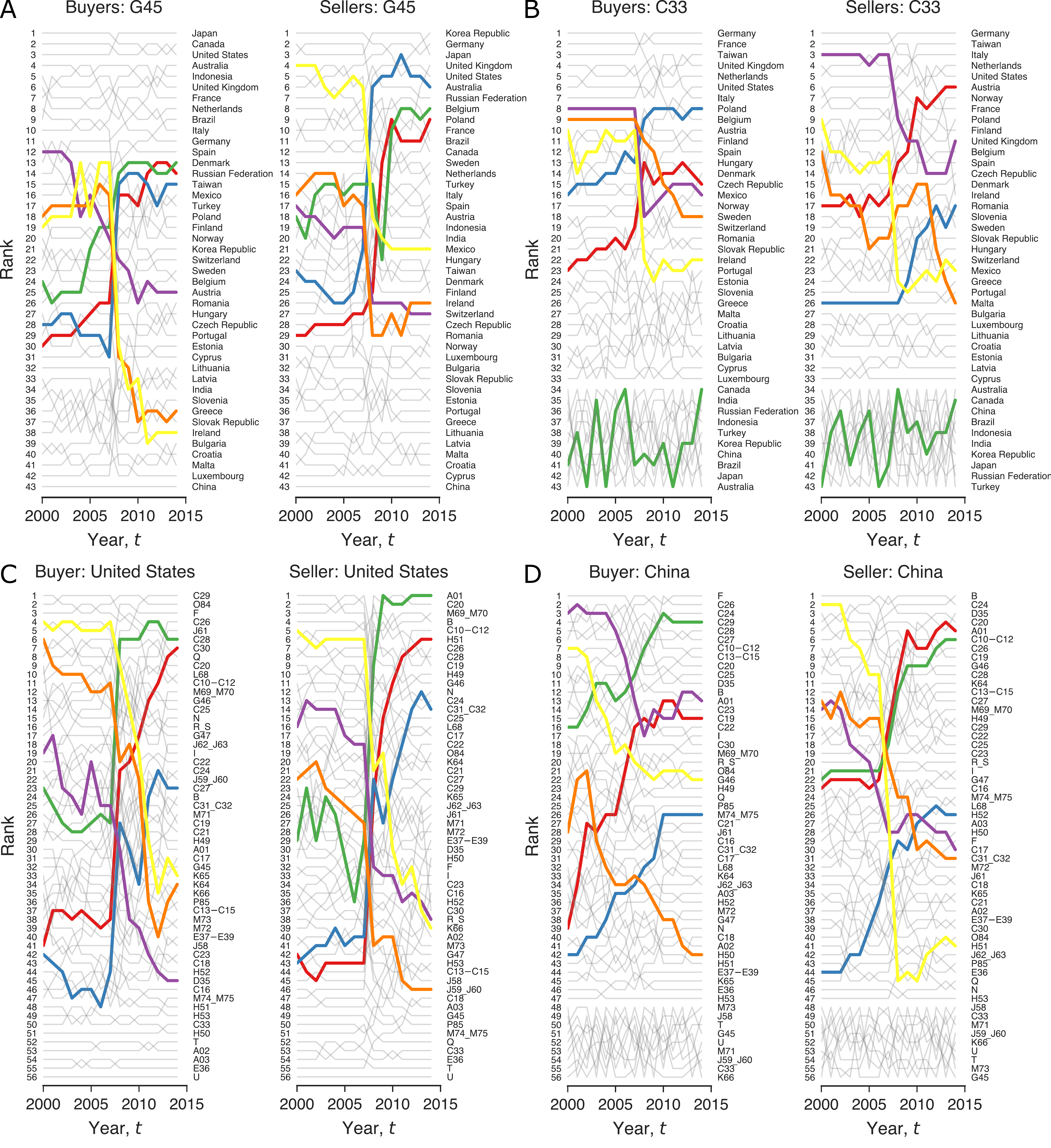}
\caption{{\bf Variation in eigenvector centrality of countries within individual industries and of industries within countries.} In each panel, the left-hand $y$-axis shows the rankings of countries (A-B) or industries (C-D) in 2000, and the right-hand $y$-axis shows the same rankings in 2015. Each panel highlights the top three countries (A-B) or industries (C-D) with the largest increase and decrease in ranking over the whole period. A) Winners and losers in the industry of wholesale and retail trade and repair of motor vehicles and motorcycles (G45). B) Winners and losers in the industry of repair and installation of machinery and equipment (C33). C) Global purchases and sales of the US. The top three purchasing industries with the largest increase in ranking were manufacture of other transport equipment (C30 - red), manufacture of electrical equipment (C27 - blue), and manufacture of machinery and equipment n.e.c. (C28 - green), whereas the purchasing industries with the largest decrease in ranking were electricity, gas, steam and air conditioning supply (D35 - purple), financial service activities, except insurance and pension funding (K64 - orange), and insurance, reinsurance and pension funding, except compulsory social security (K65 - yellow). The top three supplying industries with the largest increase in ranking were air transport (H51 - red), manufacture of furniture, other manufacturing (C31-C32 - blue), and crop and animal production, hunting and related service activities (A01 - green), whereas the three supplying industries with the largest decrease in ranking ``other'' service activities (R-S - purple), motion picture, video and television program production, sound recording and music publishing activities, programming and broadcasting activities (J59-J60 - orange), and activities auxiliary to financial services and insurance activities (K66 - yellow). D) Global purchases and sales of China. The top three purchasing industries with the largest increase in ranking were manufacture of coke and refined petroleum products (C19 - red), other professional, scientific, and technical activities, and veterinary activities (M74-M75 - blue), and manufacture of motor vehicles, trailers and semi-trailers (C29 - green), whereas the three purchasing industries with the largest decrease in ranking were crop and animal production, hunting and related service activities (A01 - purple), water transport (H50 - orange), wholesale trade, except motor vehicles and motorcycles (G46 - yellow). The top three supplying industries with the largest increase in ranking were crop and animal production, hunting and related service activities (A01 - red), warehousing and support activities for transportation (H52 - blue), other professional, scientific and technical activities, and veterinary activities (M74-M75 - green), whereas the three supplying industries with the largest decrease in ranking were manufacture of paper and paper products (C17 - purple), manufacture of furniture and other manufacturing (C31-C32 - orange), and air transport (H51 0 yellow). The full list of labels for industries can be found in Table~\ref{t:activities}.}
\label{fig:sector_rank}
\end{figure*}

We investigate the contribution of individual industries to the localization effect observed globally in the system~\cite{de2017disease}. To this end, we computed the contribution of individual layers to the IPR using the following measure:
\begin{equation}
   \text{IPR}(\alpha)=\sum_k (\theta_k^{M[\alpha]})^4,
\end{equation}
where $\theta_k^{M[\alpha]}$ is the eigenvector centrality of node $k$ in layer $\alpha$. Using this measure, we can uncover the industries that most contributed to the global localization effect, from the perspective of both buyers and sellers. Fig.~\ref{fig:ipr}C and Fig.~\ref{fig:ipr}D show the curves obtained for purchases and sales, respectively, in each layer $\alpha$ and highlight the industries characterized by a major drop of IPR in the period. The figures suggest that, for both purchases and supplies, the values of IPR($\alpha$) dropped just before the 2008 crisis. While many of the industries returned to a highly localized state (IPR$\approx 1$), other industries remained less localized and more homogeneous by distribution of power. For example, purchases in retail trade (not including motor vehicle and motorcycles) and real estate activities experienced a decay of localization during the crisis but quickly returned to a centralized power structure. By contrast, purchases within the industries of repair and installation of machinery and equipment, wholesale and retail trade and repair of motor vehicles and motorcycles, and household activities lost their market power concentration in 2007. Similar patterns can be found with suppliers. For example, administrative and support service activities experienced a decay in localization, but then slowly returned to a more centralized structure. On the other hand, other industries such as the wholesale and retail trade and repair of motor vehicles and motorcycles, activities auxiliary to financial services and insurance activities, and advertising and market research remained less centralized for the rest of the period.

We now illustrate how countries experienced variations in economic dominance within individual industries and how individual industries contributed to the rise and fall of countries. To this end, here we focus on the two industries that experienced the largest variation in the power concentration of purchases and supplies (i.e., repair and installation of machinery and equipment, and wholesale and retail trade and repair of motor vehicles and motorcycles, respectively; see Fig.~\ref{fig:ipr}C,D) and the two countries with the largest variation in economic dominance (i.e., the US and China; see Fig.~\ref{fig:centrality}C,D).  

To calculate the ranking of countries within individual industries, we computed the eigenvector centrality $\theta_{i \alpha}$ for each buyer (seller) $i$ in each industry $\alpha$ given by the eigenvector centralities of the supra-adjacency matrix (Eq.~\ref{centrality:multilayer}). We then computed the ranking of countries, for a given industry $\alpha=\alpha^*$, by sorting $\theta_{i \alpha^*}$ from the highest to the lowest value. Similarly, we computed the ranking of industries $\alpha$, for a giving country $i=i^*$, by sorting $\theta_{i^* \alpha}$ from the highest to the lowest value. The above procedure is then repeated for every year in our data set. Results for the two industries and the two countries are shown in Fig~\ref{fig:sector_rank}. It is worth noting that the countries with a dominant position in these industries (see the five top-ranked countries in Fig~\ref{fig:sector_rank})A-B or the countries with a negligible role (see the five lowest-ranked ones in Fig~\ref{fig:sector_rank}A-B) tend to maintain their positions over time. The ``market movers'' are the countries that typically occupy the middle of the ranking. Fig~\ref{fig:sector_rank}C-D sheds more light on how industries contributed to the power dynamics of the US and China and to the reversal of leading role between the two countries between 2007 and 2008. For example, the US experienced its largest loss of market dominance in the purchase of electricity, gas, steam, financial services and insurance, and in the supply of other service activities, television program production, broadcasting, music publishing and broadcasting activities. By contrast, China became a global leader by strengthening its market position in the purchase of coke, refined petroleum products, scientific and technical activities, and motor vehicles, and in the supply of crop and animal production, warehousing and support activities for transportation. Once again, most of these gains and losses in competitive advantage took place in 2007 before the crisis.

\subsection*{The role of domestic trade in localization transition}

\begin{figure*}[ht!]
\centering
\includegraphics[width=1\textwidth]{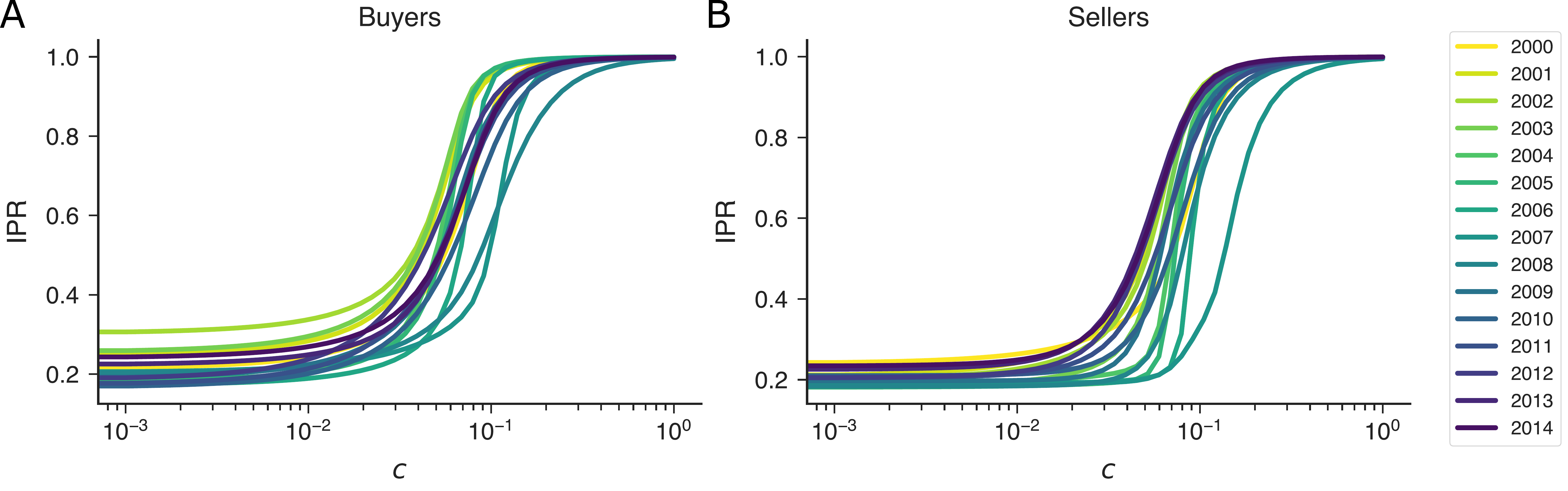}
\caption{{\bf The transition to a localized state in the multi-layer network depends on domestic trade}. By decomposing the supra-matrix into the sum of international trade and domestic trade, we use $c$ as a control parameter for varying the percentage of domestic trade accounted for when computing the IPR (Eq.~\ref{eq:decompose}). Each line shows the values of IPR calculated in a specific year and for different values of $c$. Findings suggest a transition to the localized state when $c\approx 0.08$. The panels show the localization transition for ({\bf A}) buyers  and ({\bf B}) sellers in each year from 2000 (yellow) to 2014 (purple).}
\label{fig:transition}
\end{figure*}

So far our study has suggested that localization varies across industries, and some industries are more localized than others. To further understand the sources of localization and the reversal of dominance between countries, we explore the distinct contribution to localization of international trade (i.e., the edges connecting different countries within the same layer or across different layers) and domestic trade (i.e., self-loops within layers and cross-layer connections involving the same country). To this end, we first disaggregated our supra-matrix ${\bf M}$ into an international trade matrix ${\bf T}$ and a domestic trade matrix ${\bf D}$. The domestic trade matrix includes all diagonals of each layer (i.e., the self-loops connecting a country with itself in the same layer) as well as the diagonals of the non-diagonal matrices that represent the cross-layer connections of a country with itself in different layers. We then simulated different scenarios of trade by multiplying ${\bf D}$ by a parameter $c$. This parameter $c$ therefore accounts for the role played by domestic transactions in the network. Thus, $c$ varies from $0$, when all the domestic trade is removed from the network, to $1$, where the original topology of the multi-layer network remains unchanged. We can formally define a new multi-layer network as the sum of international trade and domestic trade:
\begin{equation}
    {\bf \Tilde{M}}={\bf T}+ c\, {\bf D},
\label{eq:decompose}
\end{equation}
where $c$ is our control parameter. 

Next, for each year in the data set we calculated the IPR of ${\bf \Tilde{M}}$ by using different values of $c$ and, as usual, by distinguishing between the eigenvector centralities of buyers and sellers. Fig.~\ref{fig:transition} shows the IPR as a function of $c$. Results clearly indicate an abrupt transition to the localized regime at $c\approx 0.08$, thus suggesting that domestic trade was a main driver of localization in the network. The figure also suggests that the salience of domestic trade for localization is time-dependent. Interestingly, Fig~\ref{fig:critical_value} shows that the year-dependent critical value of the parameter $c$, at which the largest variation in IPR occurs, peaks precisely just before the financial crisis, both for buyers (Fig~\ref{fig:critical_value}A) and sellers (Fig~\ref{fig:critical_value}B). That is, at the time preceding an exogenous shock it takes a higher share of domestic trade to make the global market more heterogeneous and dominated by a minority of countries. This implies that, from the perspective of an individual country, leveraging domestic trade towards increasing or cementing the country's market dominance becomes even more critical during a crisis. 

Indeed the role played globally by domestic trade in boosting localization is reflected locally at the country's level. As shown by Fig.~\ref{fig:international}C, unlike the US and other countries, China was the only major economy to witness an uninterrupted surge in domestic trade during and beyond the financial crisis. On the other hand, the way China's international trade varied during the crisis did not differ much from what occurred in the US and Germany (see Fig.~\ref{fig:international}A,B). It is therefore the sustained and uninterrupted growth in domestic trade that enabled China to bolster its dominance of global trade, and eventually overtake the US as leading economy of the global value chains.   

\section*{Discussion}

Countries' rising market power and resilience against external shocks have become a prominent public policy issue in recent years. Here we have investigated dynamics of economic dominance using countries' eigenvector centrality in the trade multi-layer network. Our findings have uncovered a localization effect pointing to a concentration of market power on a select minority of buyers and sellers. By examining the evolution of countries' economic dominance, we uncovered two main findings. First, a reverse of market dominance between the two major economies - the US and China - took place precisely before the 2008 financial crisis. At the same time, other countries abruptly changed their global roles, and new winners and losers emerged. Second, at the global level a drop in localization took place before the crisis. In particular, the sellers' global market was the first to exhibit a significant change in power structure in 2007. These changes subsequently reverberated through the network to also affect the buyers' market.

By comparing the time series of countries' market dominance over time, we identified a number of common patterns according to which we clustered the countries. We found that similarity in power dynamics does not serve as catalyst for trade. In fact trade tends to concentrate between countries that are geographically close, yet characterized by different trends of economic dominance. Moreover, our multi-layer framework enabled us to identify the distinct contribution of individual industries to localization. While most industries managed to react to the financial crisis by retaining a heterogeneous power structure, some experienced a drop in concentration without being able to revert back to the structure they held before the crisis. This is the case, for example, of the industries of repair and installation of machinery and equipment and wholesale and retails trade and repair of motor vehicles and motorcycles.   

While policy-makers often focus on the role of exports in boosting a country's economic power in global trade, the role of domestic trade has remained largely overlooked. Our analysis of the relation between domestic trade and localization uncovered two main findings. First, domestic trade was the main driver of heterogeneity in the global power structure. Second, the salience of domestic trade for localization followed a time-dependent pattern: that is, domestic trade became most crucial for inducing power concentration in the system precisely just before and during an economic downturn. Whilst in normal times exports and, more generally, international trade enabled the US to cement its leading role, during the 2008 crisis it turned out to be domestic trade within and across industries that enabled China to overtake the US as the global market leader.

Overall, these findings have implications that straddle research and practice. Existing economic theories make divergent predictions on how the emergence of new dominant players impacts on the global political and economic landscape~\cite{Stephen2019}. There is little consensus on whether an increased unbalance in market power is likely to affect economic dynamism, income inequality, and geopolitical stability~\cite{Diez2018}. The lack of consensus partly reflects the difficulties in measuring economic power, both at the level of a firm and of a country. For example, previous macroeconomic studies have traditionally used aggregate measures such as total exports or imports to gauge the role of countries in global trade~\cite{cingolani2017countries}. However, recent studies have pointed to the inadequacy of such measures to reflect the intricacies of the underlying global value chains~\cite{antras2012,antras2013}. Here we did not perform a comparative assessment of alternative measures, but focused only on eigenvector centrality applied to the multi-layer network. Interestingly, the localization transition that the network literature has highlighted as a potential drawback of eigenvector centrality turned out to unmask properties of the power structure that would otherwise have remained hidden with other approaches. For example, our study helps to better understand how a global crisis can affect the much-debated trade war between the US and China and shift the balance of power between the two nations. The reversal in leading role that happened in 2007 suggests that exogenous events, like a financial crisis, can be turned into opportunities for global leadership. Eventually it will be the nation that most effectively responds to such events that will gain traction and emerge as a global leader. 

Recently there have been a number of attempts in the economic literature to cluster countries into meaningful blocks with distinctive preferential trade patterns within specific industries~\cite{barigozzi2011identifying,piccardi2012existence}. Here we took a different perspective on partitioning. We clustered countries according to similarity of their time series of eigenvector centrality in the global value chains, and then evaluated the obtained clusters by inferring preferential trade patterns. Our approach allowed us to complement more traditional topological approaches as we uncovered associations between countries' market positioning and preferential trading. Our partitioning suggests that countries with similar power dynamics (and thus belonging to the same block) tend to avoid trading with one another and concentrate transactions with other partners in different power-based blocks. 

Our study also contributes to the emerging literature on the structural early signals of economic downturns~\cite{saracco2015randomizing}, and more generally on the topological antecedents and consequences of shocks in economic and financial systems~\cite{squartini2013early}. We did not assess any causal relation between structural changes and exogenous events, but simply the association between the timing of a financial shock and the emergence of new winners and losers in the market. All the analysis suggests is that changes in localization can be seen as a topological signal of an upcoming global crisis and at the same time as an opportunity for countries to develop the effective strategies for safeguarding and strengthening their market positions. China serves as a case in point. The onset of the financial crisis in 2008 spurred an abrupt reversal of roles leading China to cement its status as the world’s dominant trading nation underpinned by an uninterrupted surge in domestic trade. While the role of countries in global trade has traditionally been gauged mainly based on exports and more generally international trade, this finding may suggest a change in perspective. First, assessment of countries’ market dominance needs to account not only of exports and imports but also of domestic trade taking place within the global value chains. Second, the role that domestic trade played in the geopolitical landscape in 2007 has the potential to assist governments, world leaders and policy-makers on how to help countries to strengthen their competitive advantage during critical periods marked by extraordinary and globally unfolding events, such as financial crises or pandemics.

All this raises an intriguing possibility: despite the hardships and economic losses countries suffer in the short term, it is possible that shocks to the economy turn out to be strategic inflection points. They can be the start of a sharp decline or the opportunity to rise to new heights. A comparison between the 2008 financial crisis and the current economic downturn induced by the COVID-19 pandemic is inevitable. Like the previous global systemic crisis, COVID-19 is an exogenous shock to the economy, likely to induce structural changes and a repositioning of countries in the global value chains. In an increasingly perilous global economy, the 2008 crisis can certainly offer useful insights to policy-makers and governments on how to redesign effective geo-economic strategies to chart countries’ way forward and redefine their global roles. The new emphasis on production reshoring and relocation of supply chains, the current climate crisis, the accelerating technological revolution and the fast-changing geopolitical landscape will likely set the scene for a new balance of power between countries, and the emergence of new winners and losers in a reshaped world order.

\section*{Materials and Methods}
{\bf Data:} Our study draws on data from the WIOD (Release 2016) covering $28$ EU countries and $15$ other major countries in the world within the period from 2000 to 2014. For every year, a World Input-Output Table (WIOT) is provided in current prices, expressed in millions of US dollars (USD). Each table represents economic transactions among the $56$ economic activities (industries) in each country. The core of the database is a set of harmonized supply and use tables, as well as data on international trade of goods and services. These two sets of data have been integrated into sets of inter-country WIOT. The full lists of countries and industries are provided in Supplementary Information (Table~\ref{t:countries} and Table~\ref{t:activities}).

\noindent {\bf The multi-layer network:} We define our world multi-layer network as a pair $M=(G^{M},C^{M})$, where $G^{M}=\{G^{M}_{\alpha};~\alpha~\in~\{1,\dots,k\}\}$ is a family of directed graphs $G^{M}_{\alpha}=(V^{M}_{\alpha},E^{M}_{\alpha},W^{M}_{\alpha})$ associated with the layers of $M$, and $C^{M}$ is the set of interconnections between nodes belonging to different graphs $G^{M}_{\alpha}$ and $G^{M}_{\beta}$ with $\alpha \neq \beta$. Formally, $C^{M}=\{C^{M}_{\alpha \beta};~\alpha,~\beta~\in~\{1,\dots,k\}, \alpha \neq \beta \}$ is a family of directed graphs $C^{M}_{\alpha\beta}=\{(i,j)\}$, where $\{i,j\, \in \{1,\dots, N\}\}$, $i \in V^{M}_{\alpha}$ and $j \in V^{M}_{\beta}$. We can further define the element $a_{ij}^{M[\alpha]}$ of the intra-layer adjacency matrix $A^{M[\alpha]}$ of each graph $G^{M}_{\alpha}$ as
\begin{equation}
a_{ij}^{M[\alpha]} =
  \begin{cases}
    w_{ij}^{M[\alpha]},  & \quad \text{if } (i,j) \in E^{M}_{\alpha},\\
    0,  & \quad \text{otherwise},\\
  \end{cases}\\
\end{equation}
where: $1 \leq i,j \leq N$; $1 \leq \alpha \leq k$; $c_i, c_j \in V^{M}_{\alpha}$; and $w_{c_ic_j}^{M[\alpha]}$ is the sum of the weights associated with all transactions originating from country $c_i$ within a particular industry $\alpha$ and directed to country $c_j$ within the same industry $\alpha$. Thus, an intra-layer edge between country $c_i$ and country $c_j$ in industry $\alpha$ is established when there is at least one transaction between $a_i$ and $a_j$ in $\alpha$. 

The element $a_{ij}^{M[\alpha \beta]}$ of the cross-layer adjacency matrix $A^{M[\alpha \beta]}$ corresponding to the set of interconnections $C^{M}_{\alpha \beta}$ can be defined as
\begin{equation}
a_{ij}^{M[\alpha \beta]} =
  \begin{cases}
     w_{ij}^{M[\alpha \beta]}, & \quad \text{if } (i, j) \in C^{M}_{\alpha \beta}, \\
    0,  & \quad \text{otherwise,}\\
  \end{cases}
\end{equation}
 where $1 \leq i, j \leq N$; $1 \leq \alpha, \beta \leq k$; $\alpha  \neq \beta$; $c_i \in V^{M}_{\alpha }$; $c_j \in V^{M}_{\beta}$, and $w_{ij}^{M[\alpha  \beta]}$ is the sum of the weights associated with all transactions originating from country $c_i$ within a particular industry $\alpha$ and directed towards country $i$ in industry $\beta$. Thus, a cross-layer edge between country $i$ in industry $\alpha$ and country $j$ in industry $\beta$ is established when there is at least one transaction between $a_i$ and $a_j$ across the corresponding industries. 

\noindent {\bf Dendrogram clustering:} We constructed the dendrograms using hierarchical clustering based on the correlation distance matrix obtained by 
\begin{equation}
    d=\sqrt{2 (1-\rho(\theta_i(t),\theta_j(t)))},
\end{equation}
calculated over every pair of countries, where $\theta_i(t)$ is the time series of eigenvector centralities of country $i$ and $d$ is the distance defined in the interval $[0,2]$. We further used the Ward's linkage criteria to obtain the dendrogram, as implemented in the {\it Python} package {\it Scipy}~\cite{scipy}.

To determine the number of clusters, we found the threshold distance that maximizes the silhouette score~\cite{rousseeuw1987silhouettes}. This coefficient quantifies the consistency of the clustering procedure and is defined by the average value of
\begin{equation}
s_i = \frac{b_i-a_i}{\max(a_i,b_i)}\,,
\end{equation}
where $a_i$ is the cohesion (the average intra-cluster distance) and $b_i$ is the separation (the average nearest-cluster distance) for the $i$-th country. The higher value of the silhouette coefficient represents the best cluster configuration. We used the Python module \textit{scikit-learn}~\cite{pedregosa2011scikit} to compute the silhouette scores and the \textit{SciPy}~\cite{scipy} package to compute the correlation distance matrix (Fig.~\ref{sfig:matrix}).

\noindent {\bf Null models to assess clusters}: To ascertain whether the obtained clusters differ from those obtained with more traditional network partitioning based on the tendency of countries to trade more value within than between blocks, we proceeded as follows. We constructed three null models in which the original simplex trade network maintains its topology (i.e., the in- and out degree distributions) and is otherwise randomized according to three reshuffling procedures. In the first model (Model I), for each buyer (seller), the weighted stub of each incoming (outgoing) edge is connected with the unweighted stub of an outgoing (incoming) edge chosen uniformly at random across the entire network. In this way, buyers (sellers) preserve their in-strength (out-strength) but are randomly assigned different suppliers (destination markets). In the second model (Model II), weights are reshuffled globally across the network, thus randomizing nodes' in- and out-strength. Finally, in the third model (Model III), weights are reshuffled locally across each buyer's (seller's) incoming (outgoing) edges, thus preserving each node's in-strength (out-strength) and randomizing out-strength (in-strength). For each model, $1,000$ network realizations were produced and compared to the original network. 

\noindent {\bf Synthetic multi-layer network (null model)}: Given a multi-layer network $M=(G^{M},C^{M})$, we used random edge assignment to construct $1,000$ synthetic multi-layer network realizations that preserve the in- (out-) degree, in- (out-) strength, and weight distributions of the family of observed directed graphs $G^M_\alpha$ (within-layer graphs) and $C^M_{\alpha \beta},\alpha \neq \beta$ (cross-layer graphs). In practice, we randomized the columns (incoming links pointing to buyers) or rows (outgoing links departing from sellers) by blocks of the multi-layer adjacency matrix, where each block includes intra-layer connections (i.e., diagonal matrices given by $G^M_\alpha$), or the inter-layer connections (i.e., off-diagonal matrices given by $C^M_{\alpha \beta},\alpha \neq \beta$).

\section*{Acknowledgments}
F.A.R. acknowledges CNPq (Grant No. 309266/2019-0) and FAPESP (Grants No. 2019/23293-0). Y. M. acknowledges support from the Government of Arag\'on, Spain through a grant to the group FENOL, by MINECO and FEDER funds (grant FIS2014-55867-P) and by the European Commission FET-Proactive Project Dolfins (grant 640772). 

\bibliography{references}
\clearpage
\newpage
\onecolumngrid
\section*{Supplementary Material}
\setcounter{figure}{0}
\makeatletter 
\renewcommand{\thefigure}{S\@arabic\c@figure}
\renewcommand{\thetable}{S\@arabic\c@table}

\tableofcontents

\newpage
\subsection*{Data description}

Table~\ref{t:countries} shows the list of the $43$ countries (excluding the ``Rest of the World''), and Table~\ref{t:activities} the $56$ NACE Rev.2 economic activities included in the WIOD.

\begin{table}[h]
\center
\begin{tabular}{|p{12cm}|}
\hline
\\
{\bf Country name (ISO Alpha-3 Code)} \\
\\
\hline
\\
Australia (AUS), Austria (AUT), Belgium (BEL), Bulgaria (BGR), Brazil (BRA), Canada (CAN), Switzerland (CHE), China (CHN), Cyprus (CYP), Czech Republic (CZE), Germany (DEU), Denmark (DNK), Spain (ESP), Estonia (EST), Finland (FIN), France (FRA), United Kingdom (GBR), Greece (GRC), Croatia (HRV), Hungary (HUN), Indonesia (IDN), India (IND), Ireland (IRL), Italy (ITA), Japan (JPN), Korea, Rep. (KOR), Lithuania (LTU), Luxembourg (LUX), Latvia (LVA), Mexico (MEX), Malta (MLT), Netherlands (NLD), Norway (NOR), Poland (POL), Portugal (PRT), Romania (ROU), Russian Federation (RUS), Slovak Republic (SVK), Slovenia (SVN), Sweden (SWE), Turkey (TUR), Taiwan (TWN), United States (USA)\\

\hline
\end{tabular}
\caption{\textbf{Countries in the WIOD 2016 Release}}
\label{t:countries}
\end{table}

\newpage
\begin{longtable}{|p{3cm}|p{8cm}|}

\hline \multicolumn{1}{|c|}{\textbf{NACE Rev. 2 Division}} & \multicolumn{1}{c|}{\textbf{Description of economic activities}} \\ \hline 
\endfirsthead

\multicolumn{2}{c}%
{{\bfseries \tablename\ \thetable{} -- continued from previous page}} \\
\hline \multicolumn{1}{|c|}{\textbf{NACE Rev. 2 Division}} & \multicolumn{1}{c|}{\textbf{Description of economic activity}} \\ \hline 
\endhead

\hline \multicolumn{2}{|r|}{{Continued on next page}} \\ \hline
\endfoot

\hline \hline
\endlastfoot

A01 & Crop and animal production, hunting and related service activities\\ \hline
A02    &    Forestry and logging\\ \hline
A03    &    Fishing and aquaculture\\ \hline
B    &    Mining and quarrying\\ \hline
C10-C12    &    Manufacture of food products, beverages and tobacco products\\ \hline
C13-C15    &    Manufacture of textiles, wearing apparel and leather products\\ \hline
C16    &    Manufacture of wood and of products of wood and cork, except furniture; manufacture of articles of straw and plaiting materials\\ \hline
C17    &    Manufacture of paper and paper products\\ \hline
C18    &    Printing and reproduction of recorded media\\
C19    &    Manufacture of coke and refined petroleum products \\ \hline
C20    &    Manufacture of chemicals and chemical products \\ \hline
C21    &    Manufacture of basic pharmaceutical products and pharmaceutical preparations\\ \hline
C22    &    Manufacture of rubber and plastic products\\ \hline
C23    &    Manufacture of other non-metallic mineral products\\ \hline
C24    &    Manufacture of basic metals\\ \hline
C25    &    Manufacture of fabricated metal products, except machinery and equipment\\ \hline
C26    &    Manufacture of computer, electronic and optical products\\ \hline
C27    &    Manufacture of electrical equipment\\ \hline
C28    &    Manufacture of machinery and equipment n.e.c.\\ \hline
C29    &    Manufacture of motor vehicles, trailers and semi-trailers\\ \hline
C30    &    Manufacture of other transport equipment\\ \hline
C31\_C32    &    Manufacture of furniture; other manufacturing\\ \hline
C33    &    Repair and installation of machinery and equipment\\ \hline
D35    &    Electricity, gas, steam and air conditioning supply\\ \hline
E36    &    Water collection, treatment and supply\\ \hline
E37-E39    &    Sewerage; waste collection, treatment and disposal activities; materials recovery; remediation activities and other waste management services \\ \hline
F        & Construction\\ \hline
G45    &    Wholesale and retail trade and repair of motor vehicles and motorcycles\\ \hline
G46    &    Wholesale trade, except of motor vehicles and motorcycles\\ \hline
G47    &    Retail trade, except of motor vehicles and motorcycles\\ \hline
H49    &    Land transport and transport via pipelines\\ \hline
H50    &    Water transport\\ \hline
H51    &    Air transport\\ \hline
H52    &    Warehousing and support activities for transportation\\ \hline
H53    &    Postal and courier activities\\ \hline
I    &    Accommodation and food service activities\\ \hline
J58    &    Publishing activities\\ \hline
J59\_J60    &    Motion picture, video and television program production, sound recording and music publishing activities; programming and broadcasting activities\\ \hline
J61    &    Telecommunications\\ \hline
J62\_J63    &    Computer programming, consultancy and related activities; information service activities\\ \hline
K64    &    Financial service activities, except insurance and pension funding\\ \hline
K65    &    Insurance, reinsurance and pension funding, except compulsory social security\\ \hline
K66    &    Activities auxiliary to financial services and insurance activities\\ \hline
L68    &    Real estate activities\\ \hline
M69\_M70    &    Legal and accounting activities; activities of head offices; management consultancy activities\\ \hline
M71    &    Architectural and engineering activities; technical testing and analysis\\ \hline
M72    &    Scientific research and development\\ \hline
M73    &    Advertising and market research\\ \hline
M74\_M75    &    Other professional, scientific and technical activities; veterinary activities\\ \hline
N    &    Administrative and support service activities\\ \hline
O84    &    Public administration and defense; compulsory social security\\ \hline
P85    &    Education\\ \hline
Q    &    Human health and social work activities\\ \hline
R\_S    &    Other service activities\\ \hline
T    &    Activities of households as employers; undifferentiated goods- and services-producing activities of households for own use\\ \hline
U    &    Activities of extraterritorial organizations and bodies\\ \hline

\hline
\caption{\textbf{Economic activities in WIOD 2016 Release}}
\label{t:activities}
\end{longtable}

\clearpage
\subsection*{Hierarchical structure of the multi-layer network}
\begin{figure*}[!h]
\centering
\includegraphics[width=0.99\textwidth]{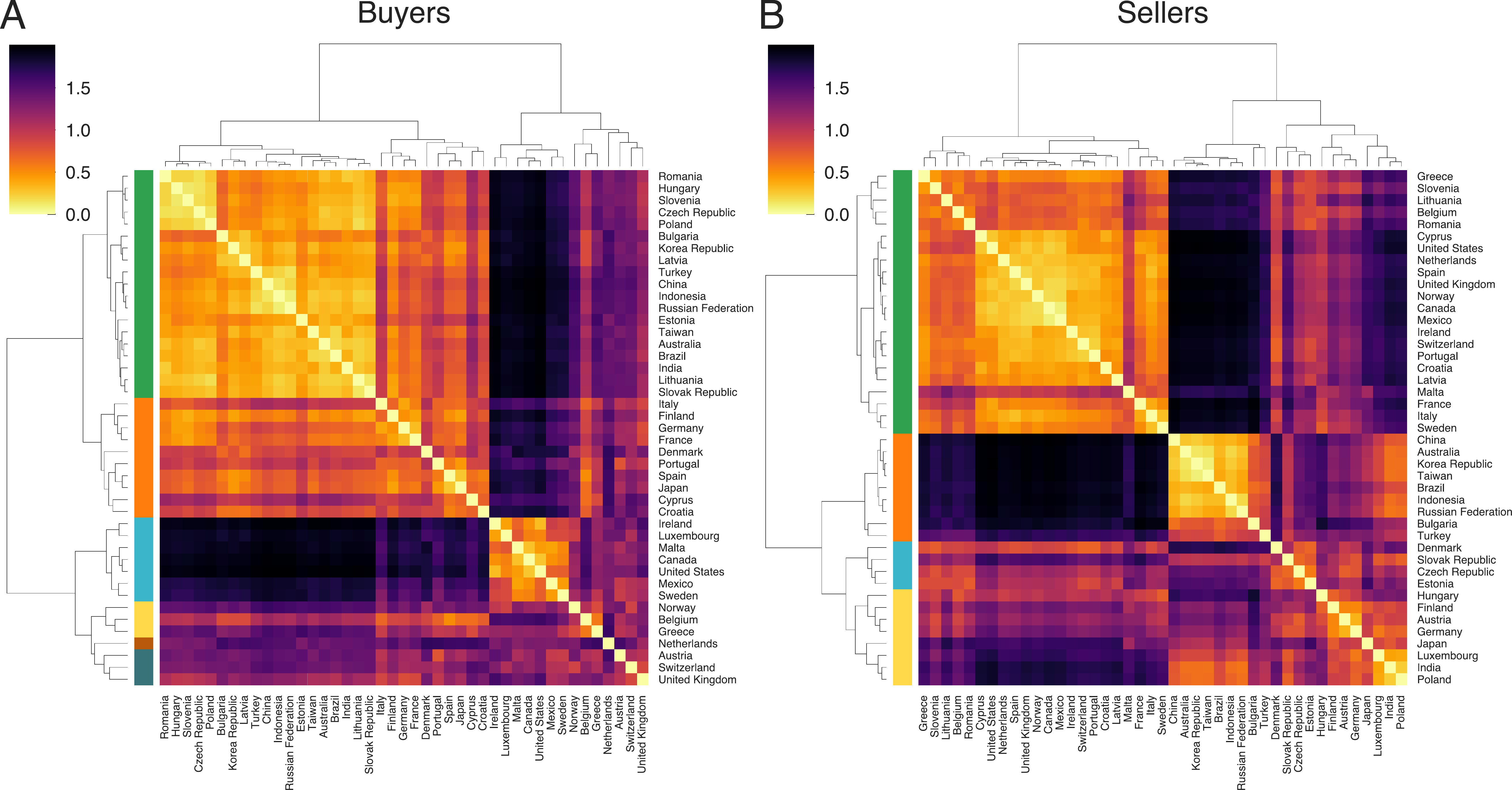}
\caption{{\bf Hierarchical structure of the multi-layer network}. Matrix plot of the correlation distance among all pairs of time series of eigenvector centrality for buyers ({\bf A}) and sellers ({\bf B}). The color of each cells is proportional to the correlation distance between the corresponding time series. The dendrograms associated with this matrix show the hierarchical clustering based on Ward's minimum variance method. The colored squares located below the dendrogram branches indicate the clusters obtained by cutting the dendrogram at the threshold distance that maximizes the silhouette score. The buyers' distance matrix has $6$ clusters, whereas the sellers' distance matrix only $4$ clusters.
}
\label{sfig:matrix}
\end{figure*}  

\newpage
\subsection*{Hierarchical (nested) stochastic block model}
\begin{figure*}[!h]
\centering
\includegraphics[width=0.9\textwidth]{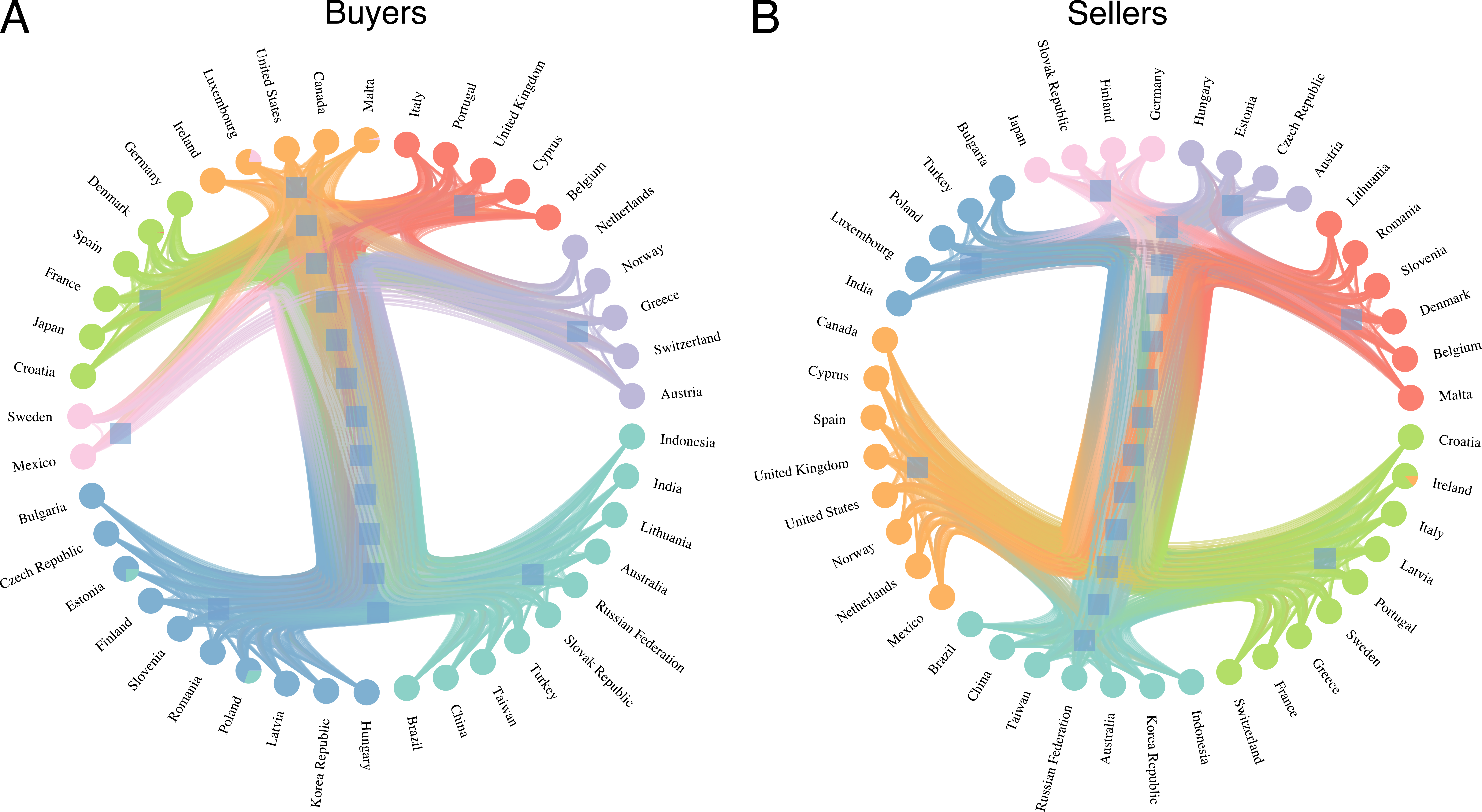}
\caption{\textbf{Hierarchical modular structure of the correlation network and marginal probabilities of node membership}. In this representation, nodes represent countries, and the pie divisions represent the marginal posterior probability that a node belongs to a given group (the different colors). The probabilities were obtained by collecting the node membership for $10,000$ sweeps of a Metropolis-Hastings acceptance-rejection Markov Chain Monte Carlo with multiple moves to sample hierarchical network partitions, at intervals of $10$ sweeps. The edges and their weights are proportional to the correlation distance of the eigenvalue time series $\theta_i(t)$ and $\theta_j(t)$ associated with country $i$ and $j$. {\bf A} Modular structure of buyers' correlation network shows a great overlap with the hierarchical cluster analysis (NMI$\approx0.7$). {\bf B} Similarly, the modular structure of sellers' correlation network shows a great overlap with the hierarchical cluster analysis (NMI$\approx0.6$).}
\label{fig:nested_SBM}
\end{figure*}

In addition to the hierarchical clustering analysis of the time series, we considered the hierarchical (nested) stochastic block model (nested SBM) to find the economic blocks~\cite{peixoto2014hierarchical}. We computed the modular/block structure of the network obtained from the correlation distance matrix of the time series $\theta_i(t)$ of eigenvector centralities. In this  network, each node is a country and the weights of links are the correlation distances $d_{ij}$ between the time series of countries $i$ and $j$. To run the nested SBM algorithm, we considered normal priors for the weight distribution and collected the partitions for $10,000$ sweeps of a Metropolis-Hastings acceptance-rejection Markov Chain Monte Carlo~\cite{peixoto2014efficient} with multiple moves to sample hierarchical network partitions, at intervals of $10$ sweeps. The block structure obtained with the hierarchical (nested) SBM for buyers and sellers are shown in Fig.~\ref{fig:nested_SBM}A and Fig.~\ref{fig:nested_SBM}B, respectively. The different colors represent the economic blocks and the adjacency edges are bundled together for a better visualization of the network hierarchical structure. 

We further estimated the marginal probabilities of node membership using the fraction of times a node is found in a given partition on our sampled partitions data. In Fig.~\ref{fig:nested_SBM}, the pie charts illustrate the marginal probabilities (fractions of occurrence on the sampled data) that a given node belongs to a partition. 

We next compared the results of the nested SBM with the results obtained with the hierarchical clustering analysis of the time series. To do so, we use the normalized mutual information (NMI) to quantify the overlap between the clusters found with the two approaches. Results suggest a great overlap between the results found with the two approaches. The NMI is $\approx0.7$ for the buyers, and $\approx0.6$ for sellers, thus highlighting the robustness of our findings. 

\clearpage
\subsection*{Fraction of purchases and sales by economic block.}
\begin{figure*}[!h]
\centering
\includegraphics[width=0.99\textwidth]{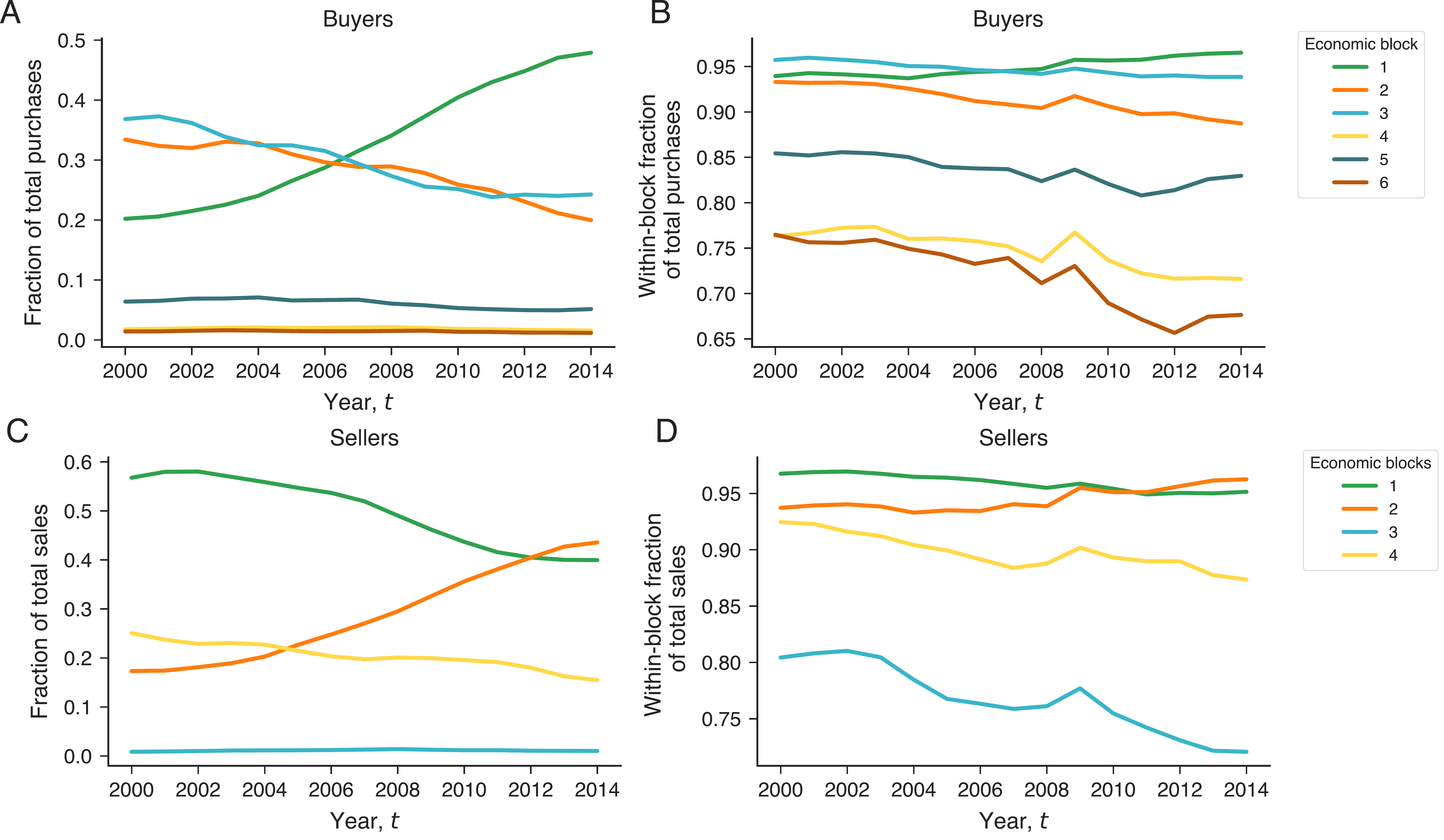}
\caption{\textbf{Fraction of purchases and sales by economic block}. In all plots, the color of each line matched the color of the corresponding economic block identified in the hierarchical clustering analysis shown in Fig.~\ref{sfig:matrix}. {\bf A} Fraction of total purchases made by members of each economic block as a function of time. The green line represents the largest block (in terms of number of countries) which in 2007 secured the largest share of purchases. {\bf B} Within-block fraction of the total purchases made by members of each economic block as a function of time. {\bf C} Fraction of total sales by members of each economic block as a function of time. The orange line represents the third largest block (in terms of number of countries) which in 2013 secured the largest share of sales. {\bf D} Within-block fraction of the total sales by members of each economic block as a function of time. Notice that, for both buyers and sellers, the purchases and sales include domestic and international trade (see Fig.~\ref{fig:blocks_fraction_international} showing purchases and sales excluding domestic trade.}
\label{fig:blocks_fraction_total}
\end{figure*}  

\clearpage
\subsection*{Fraction of imports and exports by economic block}
\begin{figure*}[!th]
\centering
\includegraphics[width=0.99\textwidth]{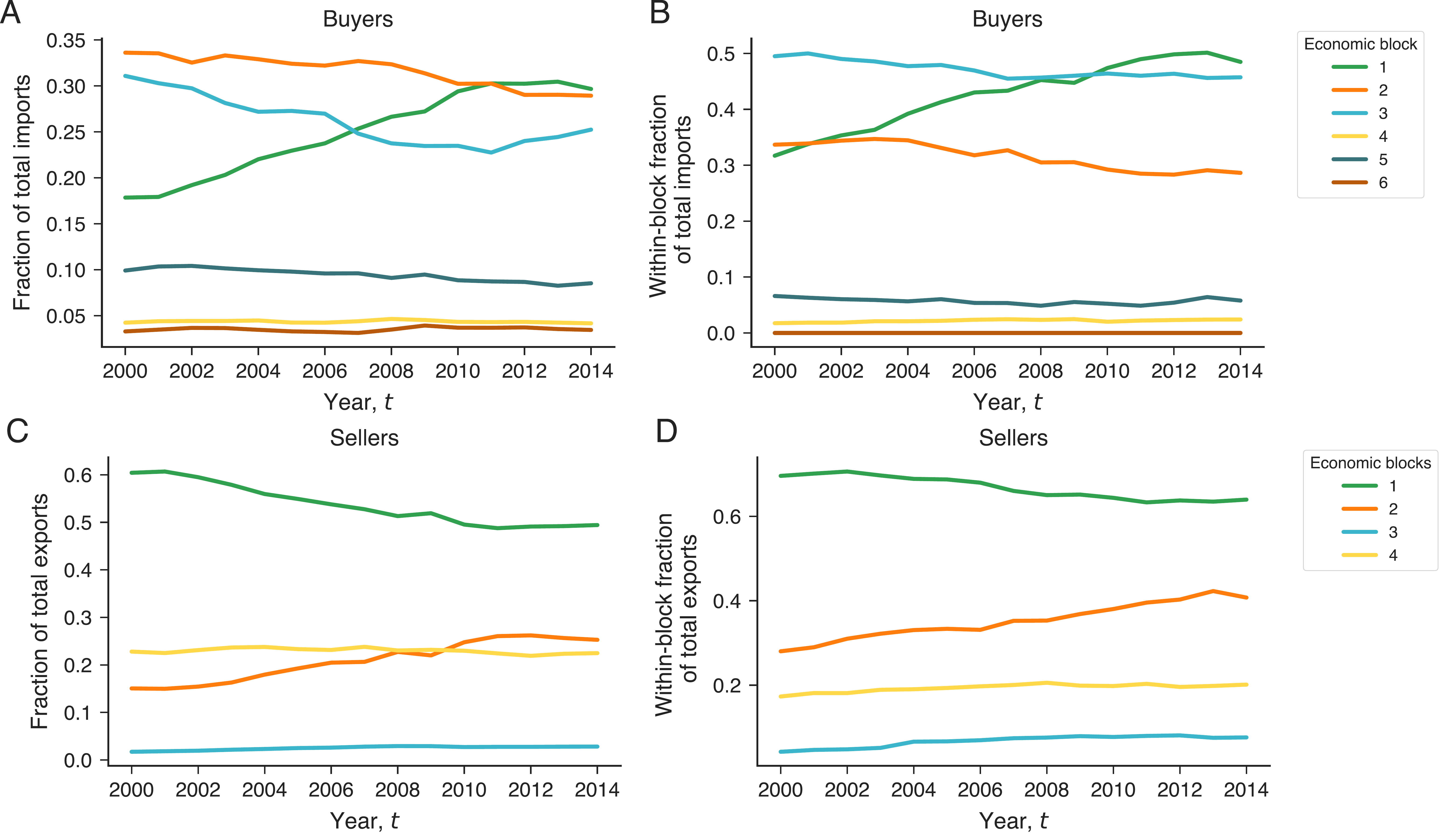}
\caption{\textbf{Fraction of imports and exports by economic block}. In all plots, the color of each line matches the color of the corresponding economic block identified in the hierarchical clustering analysis shown in Fig.~\ref{sfig:matrix}. {\bf A} Fraction of total imports by members of each economic block as a function of time. The green line represents the third largest block (in terms of number of countries) which in 2010 secured the largest share of imports. {\bf B} Within-block fraction of the total imports by members of each economic block as a function of time. {\bf C} Fraction of total exports by members of each economic block as a function of time. The orange line represents the third largest block (in terms of number of countries) which in 2009 secured the second largest share of sales. {\bf D} Within-block fraction of the total sales by members of each economic block as a function of time. }
\label{fig:blocks_fraction_international}
\end{figure*}  

\clearpage
\subsection*{Fraction of domestic purchases and sales by economic block}

\begin{figure*}[!th]
\centering
\includegraphics[width=0.99\textwidth]{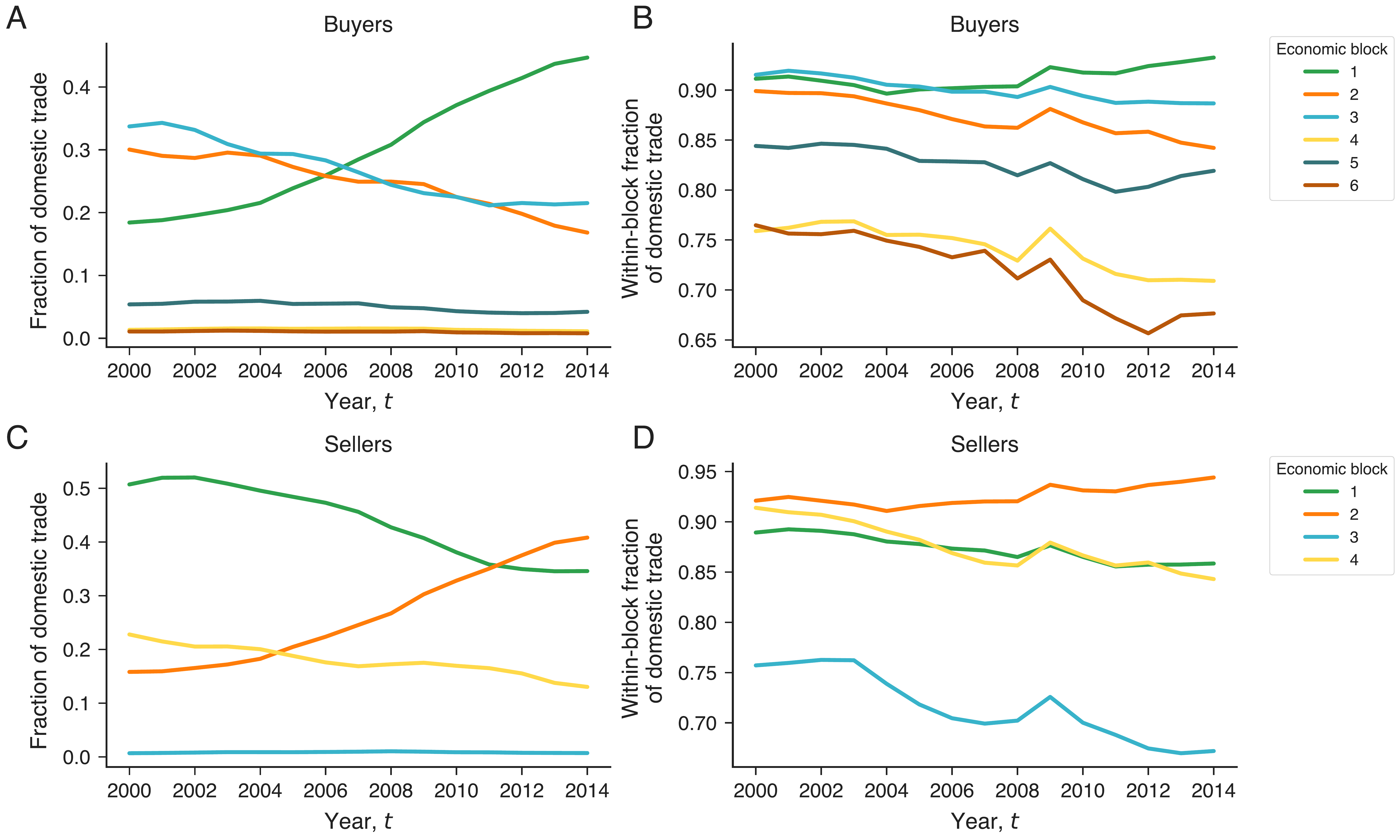}
\caption{\textbf{Fraction of domestic purchases and sales by economic block}. In all plots, the color of each line matches the color of the corresponding economic blocks identified in the hierarchical clustering analysis shown in Fig.~\ref{sfig:matrix}. {\bf A} Fraction of domestic purchases made by members of each buyer economic block as a function of time. The green line represents the third largest block (in terms of number of countries) which in 2010 secured the largest share of domestic trade. {\bf B} Within-block fraction of the domestic purchases made by members of each economic block as a function of time. {\bf C} Fraction of domestic sales by members of each economic block as a function of time. The orange line represents the third largest block (in terms of number of countries) which in 2009 secured the second largest share of sales. {\bf D} Within-block fraction of domestic sales by members of each economic block as a function of time. 
}
\label{fig:blocks_fraction_domestic}
\end{figure*}  
\clearpage
\subsection*{Null models to assess network clustering}
\begin{figure*}[!th]
\centering
\includegraphics[width=0.8\textwidth]{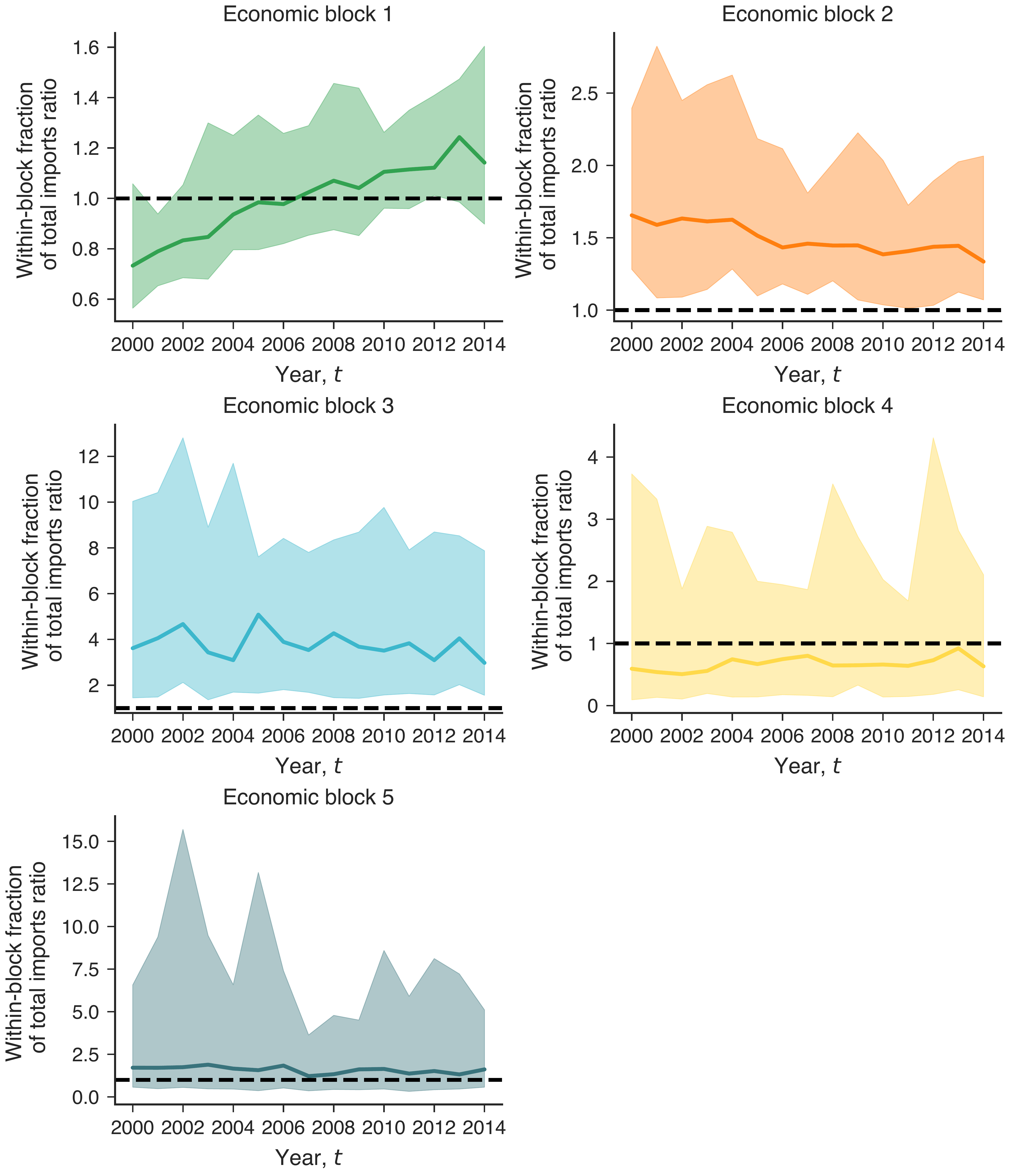}
\caption{\textbf{Null models to assess network clustering: Model I (importers)}. Within-block fraction of the total imports by members of each economic block as a function of time. In all plots, the color of each line matches the color of the corresponding economic block identified in the hierarchical clustering analysis shown in Fig.~\ref{sfig:matrix}. The shaded areas represent the 95\% confidence intervals obtained from $1,000$ realizations of the null model. We omitted the economic block no.6 since it only includes one country. }
\label{fig:importers_model1}
\end{figure*}  

\begin{figure*}[!th]
\centering
\includegraphics[width=0.8\textwidth]{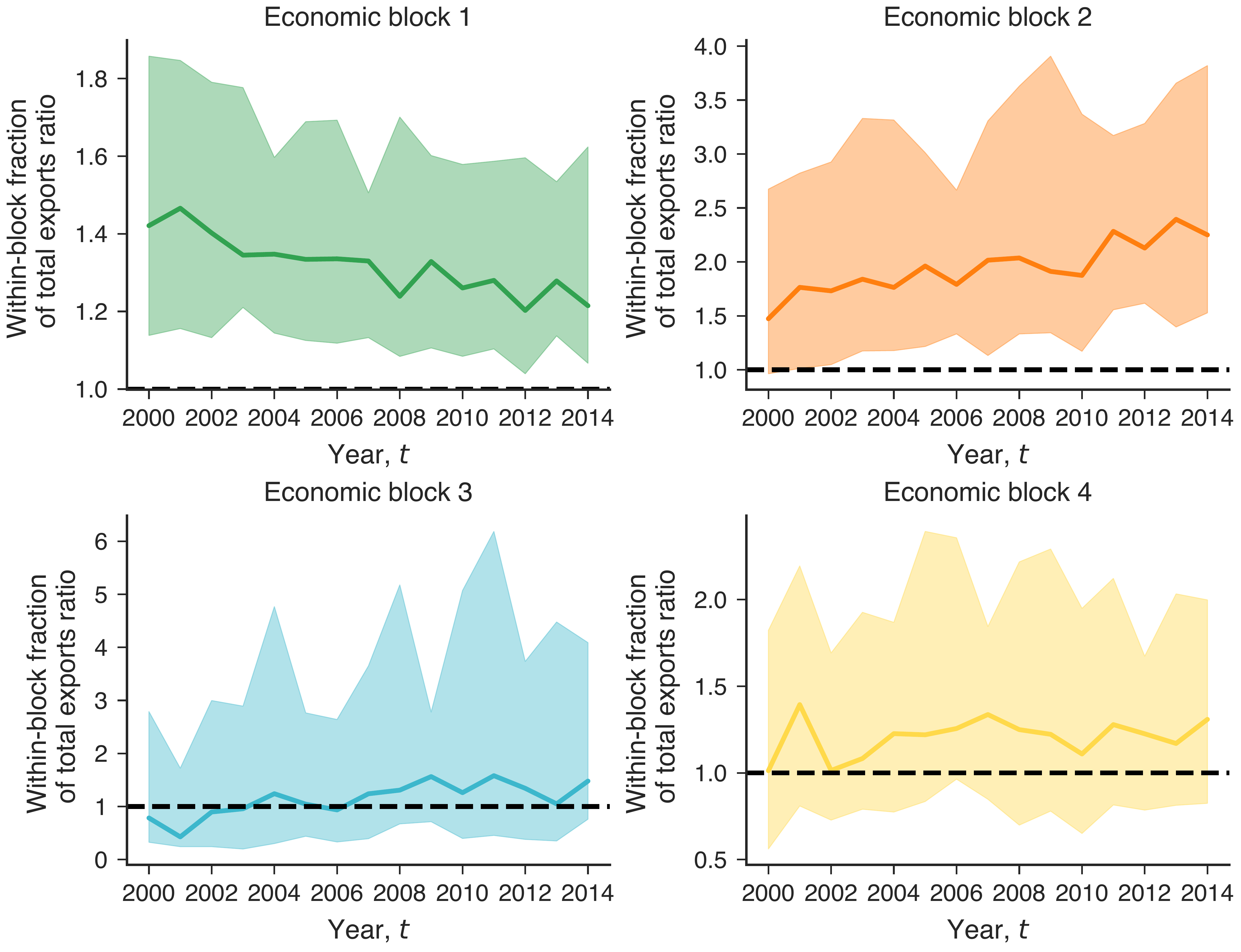}
\caption{\textbf{Null models to assess network clustering: Model I (exporters)}. Within-block fraction of the total exports by members of each economic block as a function of time. In all plots, the color of each line matches the color of the corresponding economic block identified in the hierarchical clustering analysis shown in Fig.~\ref{sfig:matrix}. The shaded areas represent the 95\% confidence intervals obtained from $1,000$ realizations of the null model. }
\label{fig:exporters_model1}
\end{figure*}

\begin{figure*}[!th]
\centering
\includegraphics[width=0.8\textwidth]{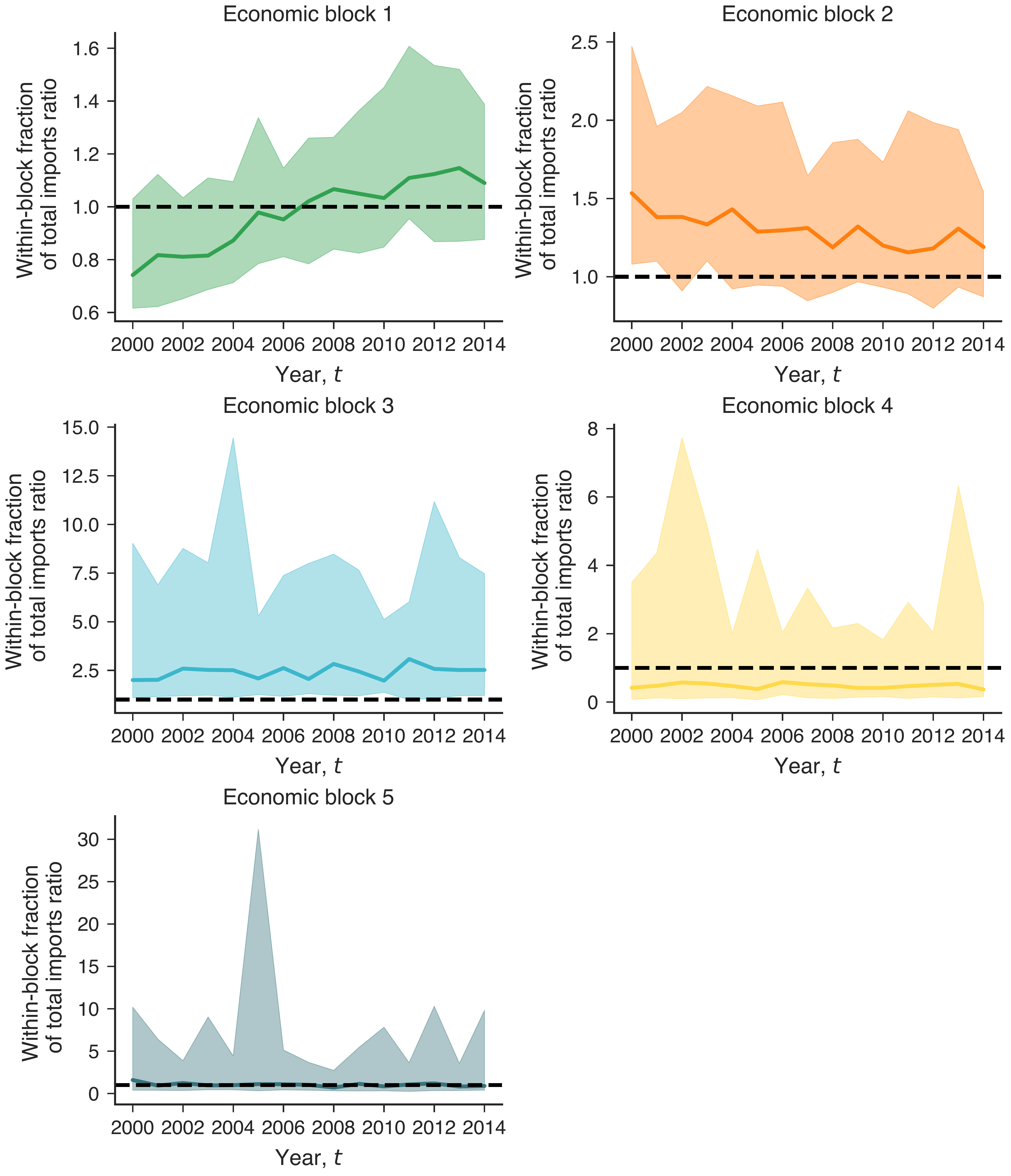}
\caption{\textbf{Null models to assess network clustering: Model II (importers)}. Within-block fraction of the total imports by members of each economic block as a function of time. In all plots, the color of each line matches the color of the corresponding economic block identified in the hierarchical clustering analysis shown in Fig.~\ref{sfig:matrix}. The shaded areas represent the 95\% confidence intervals obtained from $1,000$ realizations of the null model. We omitted the economic block no.6 since it only includes one country. }
\label{fig:importers_model2}
\end{figure*}  

\begin{figure*}[!th]
\centering
\includegraphics[width=0.8\textwidth]{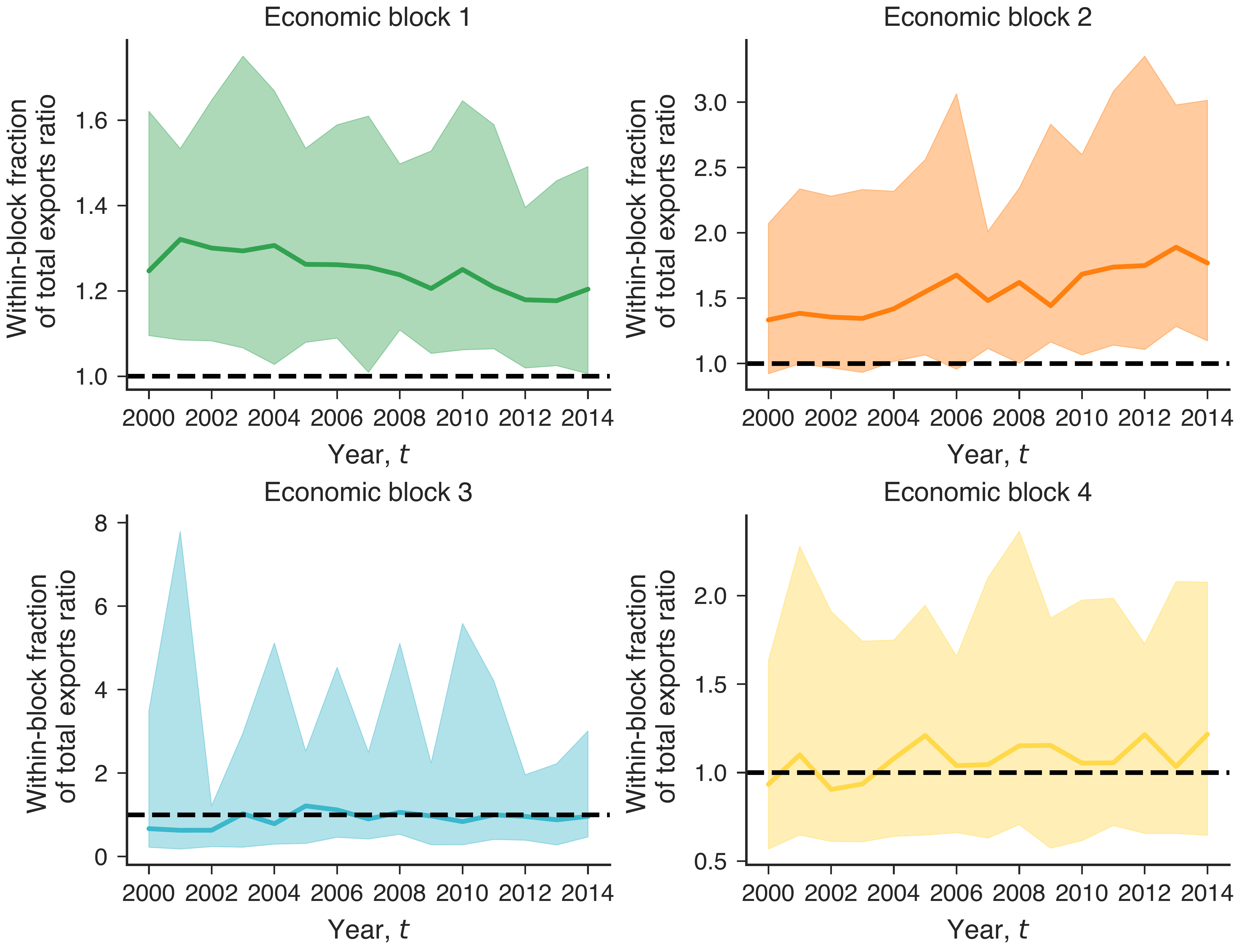}
\caption{\textbf{Null models to assess network clustering: Model II (exporters)}. Within-block fraction of the total exports by members of each economic block as a function of time. In all plots, the color of each line matches the color of the corresponding economic block identified in the hierarchical clustering analysis shown in Fig.~\ref{sfig:matrix}. The shaded areas represent the 95\% confidence intervals obtained from $1,000$ realizations of the null model.}
\label{fig:exporters_model2}
\end{figure*}  

\begin{figure*}[!th]
\centering
\includegraphics[width=0.8\textwidth]{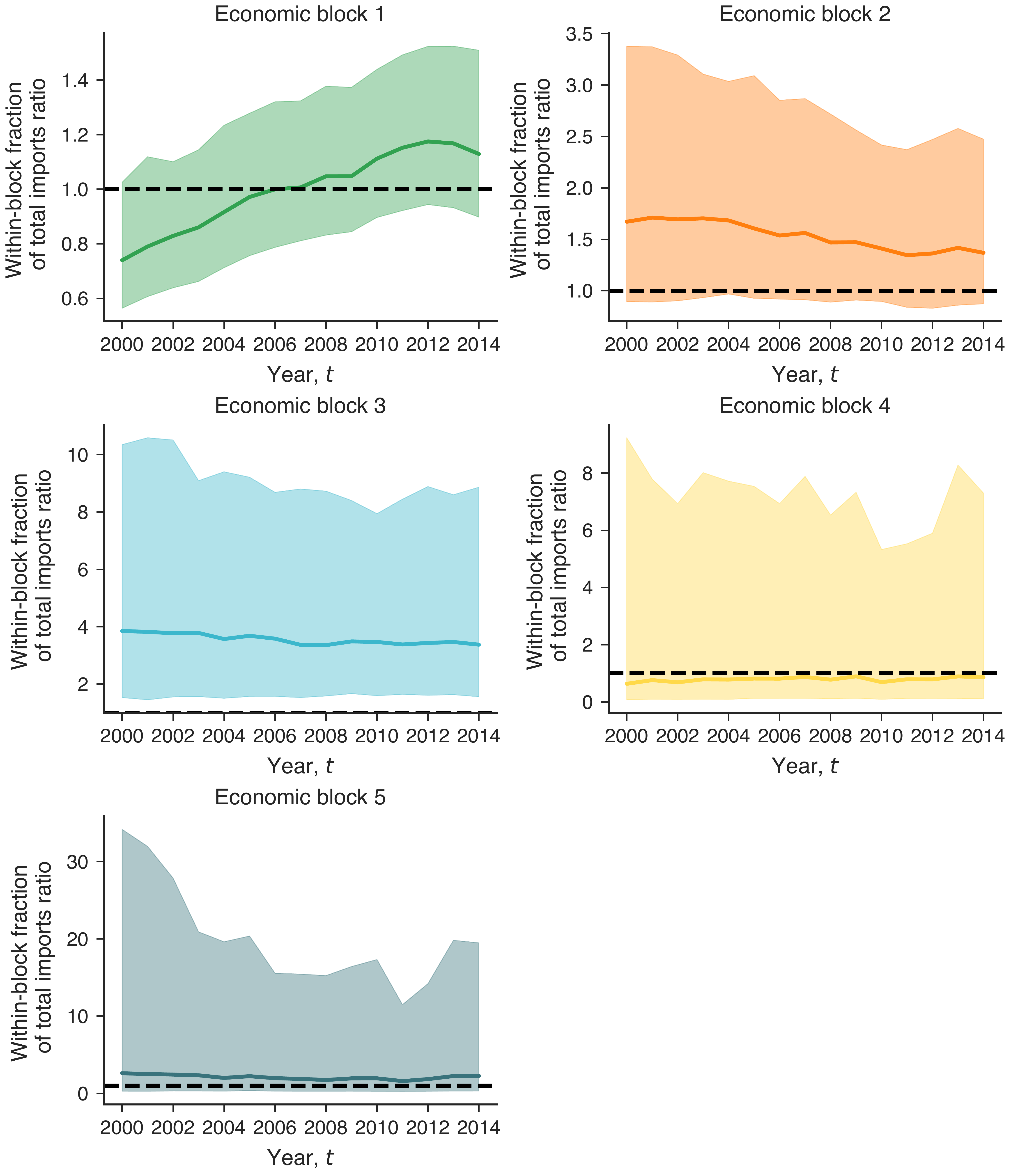}
\caption{\textbf{Null models to assess network clustering: Model III (importers)}. Within-block fraction of the total imports by members of each economic block as a function of time. In all plots, the color of each line matches the color of the corresponding economic block identified in the hierarchical clustering analysis shown in Fig.~\ref{sfig:matrix}. The shaded areas represent the 95\% confidence intervals obtained from $1,000$ realizations of the null model. We omitted the economic block no.6 since it only includes one country. }
\label{fig:importers_model3}
\end{figure*}  

\begin{figure*}[!th]
\centering
\includegraphics[width=0.8\textwidth]{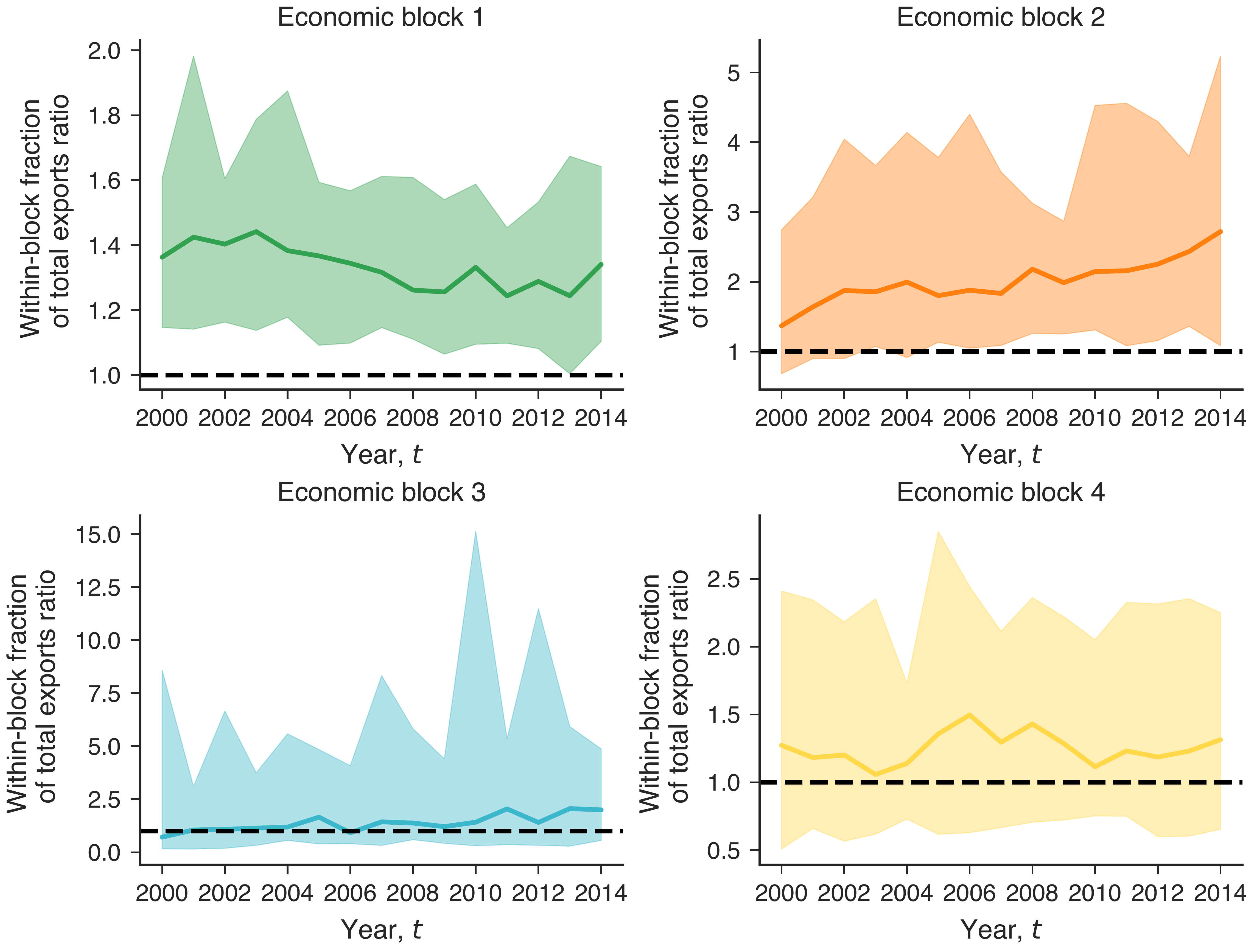}
\caption{\textbf{Null models to assess network clustering: Model III (exporters)}. Within-block fraction of the total exports by members of each economic block as a function of time. In all plots, the color of each line matches the color of the corresponding economic block identified in the hierarchical clustering analysis shown in Fig.~\ref{sfig:matrix}. The shaded areas represent the 95\% confidence intervals obtained from $1,000$ realizations of the null model. }
\label{fig:exporters_model3}
\end{figure*}  

\clearpage
\newpage
\subsection*{Geographical mapping of economic blocks}
\begin{figure*}[!th]
\centering
\includegraphics[width=0.6\textwidth]{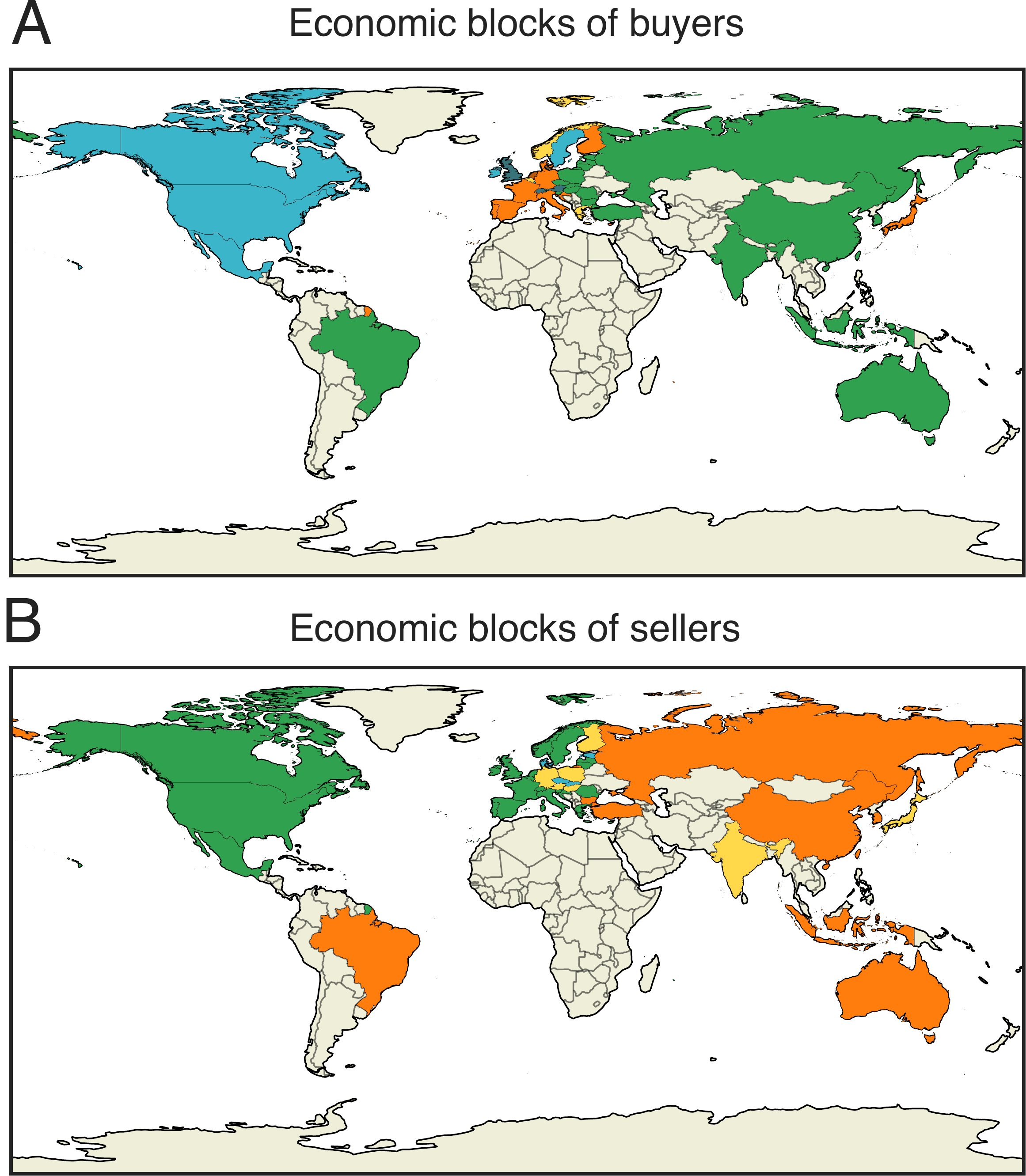}
\caption{\textbf{Geographical mapping of economic blocks}. Colors of countries match the colors of the corresponding economic blocks identified in the hierarchical clustering analysis shown in Fig.~\ref{sfig:matrix}. {\bf A} Geographical mapping of economic blocks of buyers. {\bf B} Geographical mapping of economic blocks of sellers.}
\label{fig:maps}
\end{figure*} 

\clearpage
\newpage
\subsection*{Mean geodesic distance within and between blocks}
\begin{table}[!h]
\centering
\begin{tabular}{cll}
\toprule
\textbf{Economic block} & \textbf{Within-block mean distance} & \textbf{Between-block mean distance} \\
\toprule
1 & 5630.9 & 5512.42 \\
2 & 2994.43 & 4452.62 \\
3 & 4676.4 & 5914.79 \\
4 & 1234.5 & 3661.09\\
5 & 723.47 & 3589.81\\
6 & 0 & 3397.31\\
\toprule
\end{tabular}
\caption{\textbf{Mean geodesic distance between buyers.} The first column indicates the economic blocks of buyers. The second column shows the mean geodesic distance (computed using the geographical centroid of countries) between countries within the same economic block. The third column shows the mean distance between each country within a given block and all the other countries not in the block. Notice that since block no.6 includes only one country, the mean distance is 0. Except for the largest cluster (i.e., economic block no.1), for all the other clusters of buyers within-block distances are smaller than between-block distances.} 
\label{table:distance_buyers}
\end{table}

\begin{table}[!h]
\centering
\begin{tabular}{cll}
\toprule
\textbf{Economic block} & \textbf{Within-block mean distance} & \textbf{Between-block mean distance} \\
\toprule
1 & 3129.63 & 5422.89 \\
2 & 7187.36 & 7875.84 \\
3 & 690.5 & 3523.01 \\
4 & 3489.43 & 4532.31 \\
\toprule
\end{tabular}
\caption{\textbf{Mean geodesic distance between sellers}. The first column indicates the economic blocks of sellers. The second column shows the mean geodesic distance (computed using the geographical centroid of countries) between countries within the same economic block. The third column shows the mean distance between each country within  a given block and all the other countries not in the block. For all clusters of sellers, within-block distances are smaller than between-block distances.}
\label{table:distance_sellers}
\end{table}
 
\clearpage
\newpage
\subsection*{IPR and comparison with synthetic multi-layer networks}
\begin{figure*}[!h]
\centering
\includegraphics[width=0.9\textwidth]{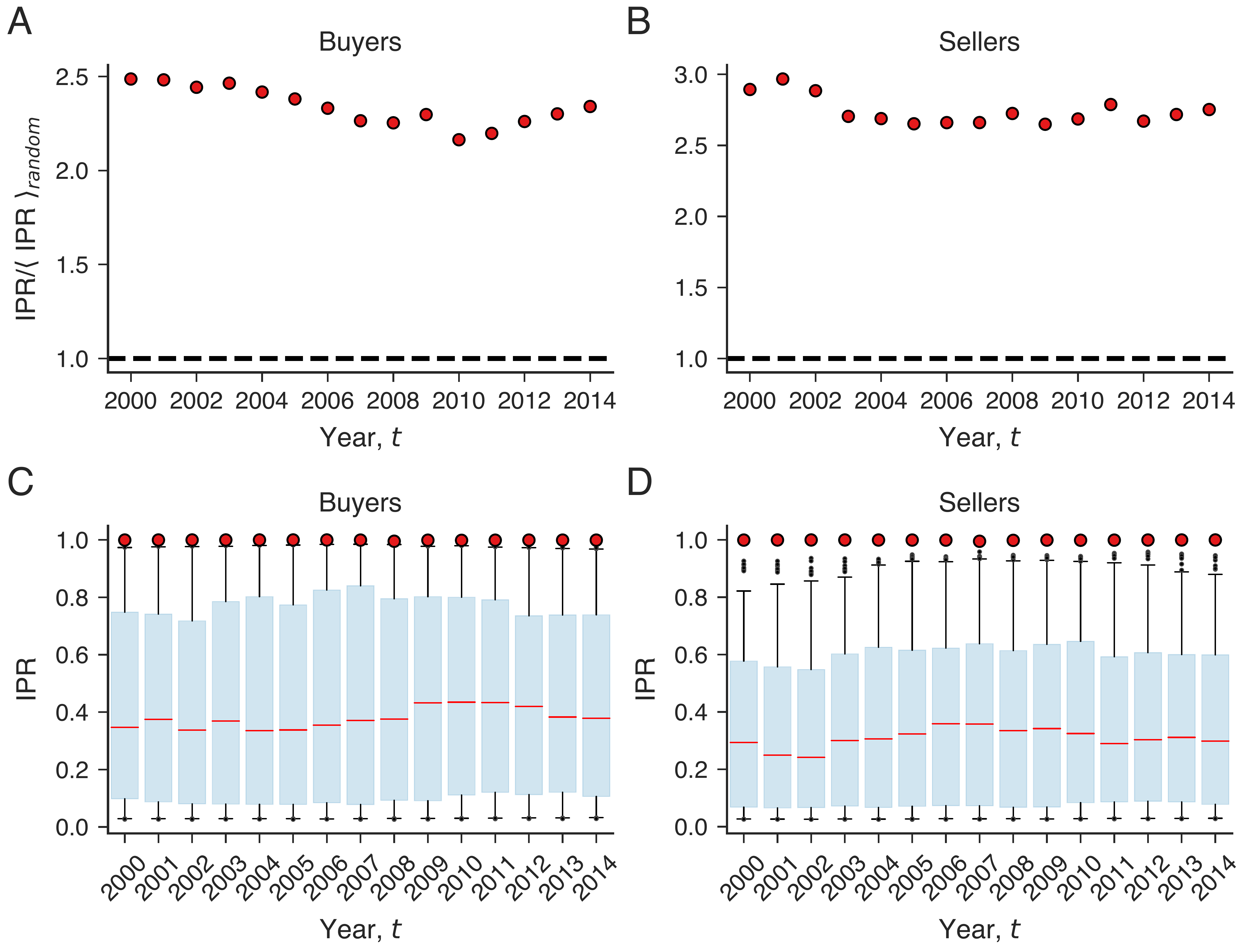}
\caption{\textbf{IPR computed on real multi-layer networks compared with the IPR obtained from synthetic random multi-layer network realizations}. {\bf A} Ratio $ \text{IPR}/\langle \text{IPR} \rangle_{\text{random}}$ for buyers (red circles). {\bf B} Ratio $ \text{IPR} /\langle \text{IPR} \rangle_{\text{random}}$ for sellers (red circles). If localization observed in the real network can be replicated using random synthetic multi-layer networks, the ratio is close to one (black dashed lines). Panels {\bf A} and {\bf B} suggest that the values of IPR, respectively for buyers and sellers, are almost three times as large as the average values found on the ensembles of random synthetic multi-layer networks. Panels {\bf C} and {\bf D} report the confidence intervals for $\text{IPR}_{\text{random}}$, based on $1,000$ realizations, respectively for buyers and sellers. All values of the observed IPR are statistically significantly different from $\text{IPR}_{\text{random}}$ at the 5\% significance level. The box-plot panels show the values from the ensembles of random network realizations for buyers ({\bf C}) and sellers ({\bf D}), and also include the values of the IPR observed in the real networks (red circles). Each blue box represents the two innermost quartiles. The whiskers represent the 95\% confidence intervals, the red horizontal lines are the medians, and the small black dots are outliers of the ensembles of synthetic random multi-layer networks.}
\label{fig:null_model}
\end{figure*}

\clearpage
\subsection*{Critical value for localization transition}
\begin{figure*}[!th]
\centering
\includegraphics[width=0.99\textwidth]{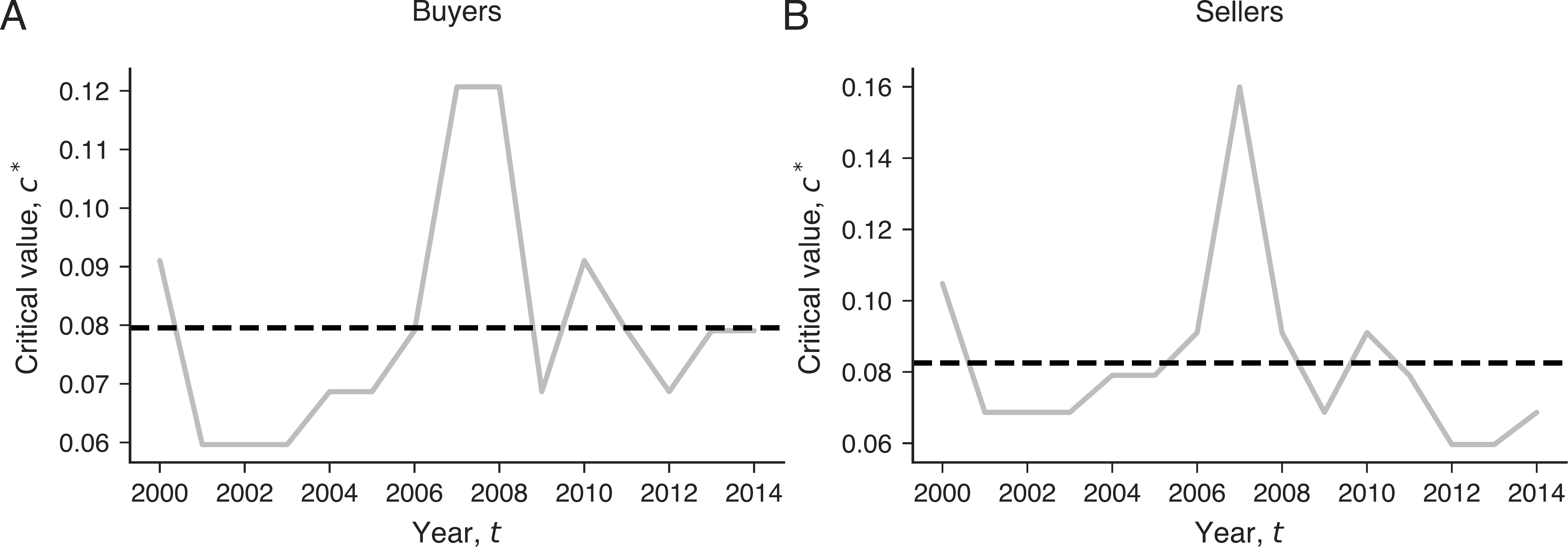}
\caption{\textbf{Critical value for localization transition}. The critical value is defined as the $c=c^*$ where the largest variation of IPR occurs. The gray lines show the variation of $c^*$ over time and the black dashed lines show the average value over the period, $\langle c^* \rangle \approx 0.08$. {\bf A} Critical values for buyers reach their maximum in 2007 and 2008. {\bf B} Critical values for sellers reach their maximum in 2007.}
\label{fig:critical_value}
\end{figure*}  

\clearpage
\subsection*{Evolution of international and domestic trade}

\begin{figure*}[!th]
\centering
\includegraphics[width=0.99\textwidth]{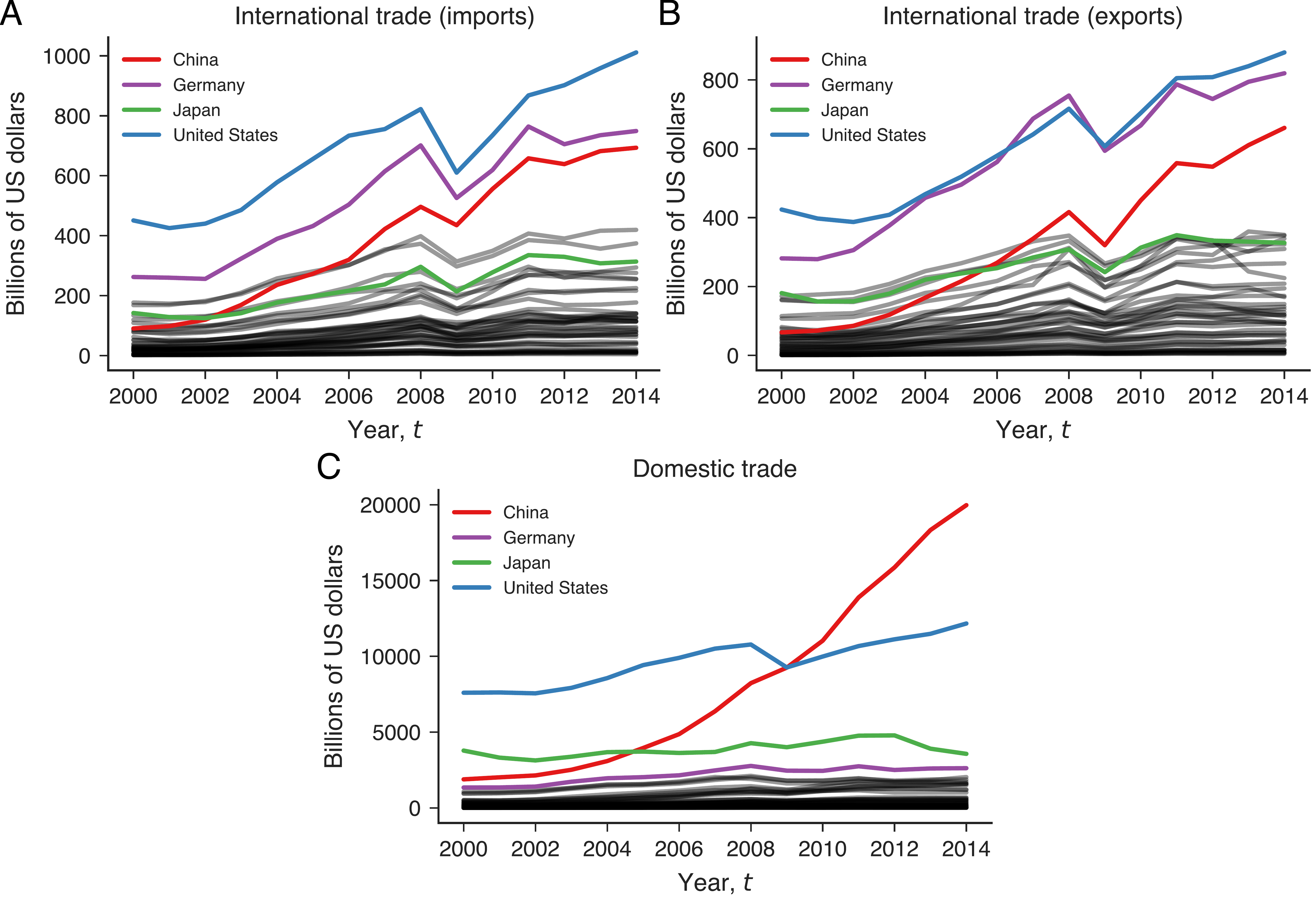}
\caption{\textbf{Evolution of countries' international trade and domestic trade}. The US has always secured the largest share of imports ({\bf A}) and exports ({\bf B}), while Germany has ranked second and in 2007 China overtook other countries reaching the third position. While Japan was characterized by a significant share of global domestic trade, its share of international trade did not rank as high as the other major countries' share. {\bf C} Domestic trade significantly increased in China during the observation period, whereas countries such the US, Germany, and Japan witnessed a much smaller growth.}
\label{fig:international}
\end{figure*}  

\end{document}